\documentclass[a4paper,11pt]{article}
\pdfoutput=1 
\usepackage{needspace}
\usepackage{jcappub} 
\usepackage{aas_macros}
\usepackage[T5]{fontenc} 
\usepackage[utf8]{inputenc}
\usepackage[dvipsnames]{xcolor}
\usepackage{mwe,bm,verbatim,cprotect,xcolor,xspace}
\usepackage{orcidlink}
\usepackage{subfig,csquotes,float, ulem}
\usepackage{bbold}
\usepackage[export]{adjustbox}
\graphicspath{{figs/}}
\definecolor{RevisionColor}{rgb}{0,0,0} 
\newcommand{\revchange}[1]{{\color{RevisionColor}#1}}

\usepackage{fontawesome5}
\makeatletter
\newcommand{\github}[1]{%
   \href{#1}{\faGithub}%
}

\usepackage[nameinlink,noabbrev]{cleveref}
\crefname{equation}{Eq.}{Eqs.}
\crefname{section}{Sec.}{Secs.}
\crefname{figure}{Fig.}{Figs.}
\crefname{table}{Tab.}{Tabs.}
\crefname{appendix}{App.}{Apps.}
\Crefname{equation}{Eq.}{Eqs.}
\Crefname{section}{Sec.}{Secs.}
\Crefname{figure}{Fig.}{Figs.}
\Crefname{table}{Tab.}{Tabs.}
\Crefname{appendix}{App.}{Apps.}

\def\be{\begin{equation}}
\def\ee{\end{equation}}

\def\ba#1\ea{\begin{align*}#1\end{align*}}
\def\bg#1\eg{\begin{gather}#1\end{gather}}

\newcommand{\code}[1]{{\texttt{#1}}}
\def\knl{k_{\mathrm{NL}}}
\renewcommand{\emph}[1]{\textit{#1}}
\newcommand{\Om}{\Omega_m}
\def\Mpch{\,h^{-1}\,\mathrm{Mpc}}
\def\iMpch{\,h\,\mathrm{Mpc}^{-1}}

\def\kmax{k_{\mathrm{max}}}
\newcommand{\dd}{\mathrm{d}}
\DeclareMathOperator{\Cov}{Cov}
\newcommand{\se}{\ensuremath{\sigma_{8}}\xspace}

\newcommand{\G}{\mathcal{G}}

\makeatletter
\renewcommand*{\thesection}{\arabic{section}}

\renewcommand*{\p@subsection}{}
\makeatother

\newcommand{\figurespath}{figs}

\newcommand{\PFS}{\mathrm{PFS}}

\newcommand{\Mnu}{M_\nu}
\newcommand{\ctilde}{\tilde{c}}
\newcommand{\Pshot}{P_{\mathrm{shot}}}
\newcommand{\rsv}{r_{\sigma_v}}
\newcommand{\nbar}{\bar{n}}

\title{Multi-tracers, multi-surveys: a joint Fisher analysis of DESI+PFS}
\author[a,b,c]{Nhat-Minh Nguyen (Nguyễn Nhật Minh)\orcidlink{0000-0002-2542-7233}}
\affiliation[a]{Center for Data-Driven Discovery, Kavli IPMU (WPI), UTIAS, The University of Tokyo, Kashiwa, Chiba 277-8583, Japan}
\affiliation[b]{Kavli IPMU (WPI), UTIAS, The University of Tokyo, 5-1-5 Kashiwanoha, Kashiwa, Chiba 277-8583, Japan}
\affiliation[c]{Institute For Interdisciplinary Research in Science and Education, ICISE, Quy Nhon, 55121, Vietnam}
\emailAdd{nhat.minh.nguyen@ipmu.jp}
\keywords{galaxy surveys, redshift surveys, galaxy clustering, power spectrum, cosmological parameters from LSS, neutrino masses from cosmology}

\abstract{%
\emergencystretch=1em
\revchange{Marginalizing over 11} effective-field-theory (EFT) nuisance parameters per tracer per redshift bin \revchange{degrades cosmological precision in one-loop analyses of galaxy power-spectrum multipoles in redshift space.} Simulation-based priors (SBP) tighten these parameters but rely on N-body simulations and halo-occupation-distribution (HOD) models. \revchange{We consider a multi-tracer Fisher analysis as a complementary approach: auto- and cross-power spectra jointly constrain cosmology, galaxy bias and stochasticity} through two channels---within a survey and across overlapping surveys---combined in a volume-partitioned joint Fisher \revchange{analysis}. We forecast across the $14{,}000\;\mathrm{deg}^2$ Dark Energy Spectroscopic Instrument (DESI) footprint, including the $\sim\!1{,}200\;\mathrm{deg}^2$ Prime Focus Spectrograph (PFS) overlap at $z\in[0.6,1.6]$ with up to 4 tracers (PFS-ELG, DESI-ELG, DESI-LRG, DESI-QSO). \revchange{We vary a common $\ln(10^{10}A_s)$, $M_\nu$ and $\Omega_m$, holding $h$, $\omega_b$ and $n_s$ fixed.} \revchange{Adding internal-DESI cross-spectra to DESI auto-spectra reduces the errors on the derived $f\sigma_8(z=1.1)$, $M_\nu$ and $\Omega_m$ by 7.1\%, 3.5\% and 3.8\%, respectively,} at $k_{\mathrm{max}}=0.20\,h\,\mathrm{Mpc}^{-1}$. \revchange{Adding PFS auto- and cross-spectra to this complete DESI multi-tracer analysis} further tightens these \revchange{errors by 3.8\%, 2.6\% and 3.2\%,} respectively, after marginalizing over residual cross-population stochasticity. \revchange{The clustering likelihood alone constrains the linear and quadratic biases and the shot noise. For these parameters it provides a complementary test of inference using SBP, while the counterterms remain weakly constrained.} The framework extends to any number of overlapping spectroscopic surveys.
Code is available at \href{https://github.com/MinhMPA/PFSxDESI-multi-fish}{\faGithub\ MinhMPA/PFSxDESI-multi-fish}.
}

\begin{document}
\date{\today}
\begin{flushright}
IPMU26-0019
\end{flushright}
\maketitle
\flushbottom

\section{Introduction}
\label{sec:intro}

Full-shape analyses of the galaxy power
spectrum, built on the effective field theory of large-scale
structure (EFTofLSS)~\cite{Baumann:2010tm,Carrasco:2012cv,Carroll:2013oxa,Senatore:2014eva}, are now standard in spectroscopic-survey clustering.
From their first applications on the Sloan Digital Sky Survey BOSS galaxy samples~\citep[e.g.,][]{DAmico:2019fhj,Ivanov:2019pdj}, full-shape EFT analyses have improved large-scale structure constraints on, for example, neutrino masses~\cite{Colas:2019ret,Ivanov:2019hqk,Chudaykin:2025lww}, primordial
non-Gaussianity~\cite{Cabass:2022wjy,DAmico:2022gki,Chudaykin:2025vdh}, early dark energy~\cite{Ivanov:2020ril,DAmico:2020ods,Smith:2020rxx}, clustering dark energy~\cite{DAmico:2020tty}, interacting dark energy~\cite{Tsedrik:2025cwc}, ultra-light axions~\cite{Rogers:2023ezo} and dark-sector
interactions~\cite{He:2023oke,He:2025npy}.
The DESI collaboration applied this framework to the DESI~DR1 full-shape clustering analyses, reporting evidence for dynamical dark energy~\cite{DESI:2024jxi,DESI:2024hhd}. \revchange{Independent reanalyses by Chudaykin et al. include} $\Lambda$CDM constraints from the joint power spectrum and bispectrum~\cite{Chudaykin:2025aux}, dark energy and neutrino masses~\cite{Chudaykin:2025lww}, and primordial non-Gaussianity~\cite{Chudaykin:2025vdh}. Combining DR1 with CMB data, Ivanov et al.~\cite{Ivanov:2026dvl} reported a robust preference for the normal neutrino mass hierarchy over the inverted hierarchy, and Chudaykin et al.~\cite{Chudaykin:2026nls} folded simulation-based priors into the same pipeline.

At next-to-leading (i.e.\ one-loop) order, the EFT framework
parameterizes our ignorance of small-scale physics \revchange{in the galaxy
power-spectrum multipoles in redshift space through bias coefficients,
counterterms, and stochastic amplitudes~\cite{Chudaykin:2020aoj,Desjacques:2018pfv}.
In this forecast, we vary 11 such parameters per tracer per redshift bin
(\cref{tab:eft_params}), with additional stochastic parameters for each included cross-spectrum.}
\revchange{Marginalizing over these parameters degrades the cosmological precision
relative to perfect nuisance knowledge~\cite{Chudaykin:2024wlw}.}

Tightening the nuisance priors is therefore a direct route to sharper
cosmological constraints. \revchange{Recent approaches to informative EFT
priors use simulations: Halo-occupation-distribution (HOD)} mock catalogs are first generated
from N-body simulations, the EFT parameters are then measured from the
mocks, either at the power-spectrum level or at the field
level, and the resulting marginal distributions of EFT parameters are used as priors in the EFT full-shape analyses~\cite{Ivanov:2024hgq,Ivanov:2024xgb,Zhang:2024thl}.
Zhang et al.~\cite{DESI:2025wzd} applied power-spectrum-level priors to DESI DR1, improving $\sigma(\se)$ by $\sim\!23\%$\revchange{, while
Chudaykin et al.}~\cite{Chudaykin:2026nls} achieved
$\sim\!50\%$ with field-level priors.
These gains are substantial, but the simulation-based priors (SBP) are only as accurate as the HOD model used to produce them~\cite{Akitsu:2024lyt}.
Notably, the SBP analysis in Chudaykin et al.~\cite{Chudaykin:2026nls} shifts $\se$ downward by $\sim\!2\sigma$ relative to conservative
priors, deepening the tension with CMB
measurements of $S_8$.
\revchange{App.~D of their updated analysis shows that this downward shift is sensitive to the quasar HOD priors.}

\revchange{A multi-tracer analysis provides a complementary route to constraining the same bias
parameters together with cosmology. These constraints come from the
clustering likelihood itself. When the samples and analysis assumptions
are matched, they can cross-check SBP-based cosmological constraints,
which rely on an external HOD-derived prior.
Here, multi-tracer analysis refers to a joint analysis of the auto- and cross-spectra of galaxy and quasar samples used as tracers of
the same underlying matter density field.}

Multi-tracer Fisher forecasts have established parameter-degeneracy breaking beyond the single-tracer reach. \revchange{Alarcon et al.}~\cite{Alarcon:2016bkr} forecasted multi-tracer cosmological constraints from overlapping spectroscopic surveys at the linear-bias level. Mergulhão et al.~\cite{Mergulhao:2021kip,Mergulhao:2023zso} extended the multi-tracer formalism to the EFTofLSS at one loop, first in real space and then in redshift space with realistic tracers.
Rubira and Conteddu~\cite{Rubira:2025scu} pushed beyond linear theory and demonstrated the joint multi-tracer constraint of bias and stochastic parameters, and Qin et al.~\cite{Qin:2025nkk} applied a joint J-PAS\,$\times$\,PFS Fisher to dark energy and neutrino-mass forecasts.
Zhao et al.~\cite{Zhao:2024eboss} realized this with eBOSS DR16, jointly analyzing LRGs and ELGs and improving $\sigma(\se)$ by 18--27\%, with the spread set by the treatment of LRG--ELG cross-stochasticity.

\revchange{In this paper, we forecast joint cosmological and nuisance-parameter
constraints from two channels of multi-tracer information: correlations
within DESI and between DESI and PFS. Building on the nonlinear multi-tracer
studies discussed above, we ask how much information PFS adds after the
internal DESI multi-tracer information has been included.}

\revchange{Using matched redshift bins and scale cuts, we isolate the information added by DESI cross-spectra,
PFS auto-spectra and DESI\,$\times$\,PFS cross-spectra.}
The internal-DESI channel covers $z\in[0.4,2.1]$ with three tracers
(LRG, ELG, QSO) and the full \revchange{$14{,}000\;\mathrm{deg}^2$}
DESI footprint. \revchange{The smaller DESI\,$\times$\,PFS overlap adds
PFS-ELG over $1{,}200\;\mathrm{deg}^2$ at $z\in[0.6,1.6]$,}
with selection and shot noise distinct from DESI-ELG\revchange{.
The joint Fisher analysis retains DESI outside the overlap.}

\revchange{We vary
$(\ln(10^{10}A_s),\Mnu,\Om)$ while holding $h,\omega_b,n_s$ fixed.
The reported cosmological constraints marginalize over the EFT nuisance
parameters, including residual cross-stochasticity, with $f\se(z)$
derived from the same cosmology.}

\revchange{Relative to the DESI auto-spectra,
including the DESI cross-spectra reduces the marginalized uncertainties
in $f\se(z=1.1)$, $\Mnu$ and $\Omega_m$ by
$7.1\%$, $3.5\%$ and $3.8\%$, respectively, at a common observed scale cut
$k_{\mathrm{max}}=0.20\,h_{\mathrm{ref}}\,\mathrm{Mpc}^{-1}$.
For the assumed PFS-ELG sample,
its auto- and cross-spectra reduce the same uncertainties by
$3.8\%$, $2.6\%$ and $3.2\%$, respectively, relative to the
DESI multi-tracer analysis.}

This paper is organized as follows. \Cref{sec:method} describes the joint multi-tracer, multi-survey Fisher framework\revchange{, as well as the four cases being compared.
\Cref{sec:fisher} specifies the survey inputs, the one-loop model, its priors and the numerical calculation.
Some readers might want to skip \cref{sec:eft_model,sec:eft_params,sec:eft_fiducials} on a first reading and optionally return to them later for details about the EFT modeling, parametrization and parameter fiducial values.}
\Cref{sec:results} presents the \revchange{nuisance and cosmological constraints, the cost of
nuisance marginalization and the robustness of the gains}.
We discuss implications in \cref{sec:discussion} and conclude in
\cref{sec:conclusions}.

\section{Joint multi-tracer, multi-survey Fisher analysis}
\label{sec:method}

\revchange{To quantify how multi-tracer information can reduce the cost of nuisance
marginalization, we first describe how auto- and cross-spectra constrain nuisance parameters,
how their Fisher information is combined across overlapping survey
volumes, and which cases we compare.}

\subsection{Marginalization cost}
\label{sec:marg_cost}

\revchange{\emergencystretch=1em
The one-loop EFTofLSS model for the redshift-space galaxy power-spectrum
multipoles $P_\ell(k)$ of order $\ell$ involves 11 nuisance parameters per tracer
per redshift bin: four bias coefficients
$b_1,b_2,b_{\G_2},b_{\Gamma_3}$~\cite{Desjacques:2016bnm}, four counterterms
$c_0,c_2,c_4,\ctilde$ that absorb unresolved small-scale physics, and
three auto-spectrum stochastic coefficients $\Pshot,a_0,a_2$ that
parameterize departures from Poisson shot noise.
All are varied explicitly and marginalized jointly in the Fisher analysis.
The amplitude-scaled combinations used as nuisance coordinates and their
prior widths are specified in \cref{sec:eft_params,tab:eft_params}.
Each included cross-spectrum introduces two additional stochastic
parameters, as described below.\par}

\revchange{We ask how much the joint auto- and cross-spectrum likelihood reduces
this marginalization cost at fixed nuisance priors.}

\Needspace{5\baselineskip}
\subsection{Cross-power constraints on nuisance parameters}
\label{sec:cross_power}

\revchange{Two channels constrain nuisance parameters jointly with cosmology}
in the present analysis. The internal-DESI channel uses
LRG\,$\times$\,ELG\,$\times$\,QSO cross-spectra over the full DESI
footprint\revchange{. The DESI\,$\times$\,PFS} overlap channel adds PFS-ELG
cross-spectra in the $1{,}200\;\mathrm{deg}^2$ footprint\revchange{.
The approach exploits the common matter fluctuations to reduce cosmic
variance in relative-bias measurements, as} developed by
Seljak~\cite{Seljak:2008xr} and McDonald \revchange{and}
Seljak~\cite{McDonald:2008sh}, extended to the EFTofLSS
multi-tracer setting by Mergulh\~ao et al.~\cite{Mergulhao:2021kip,Mergulhao:2023zso},
and applied to the joint constraint of bias and stochastic parameters by
Rubira \revchange{and Conteddu~\cite{Rubira:2025scu}.}
\revchange{For a tracer population $X$, we write the redshift-space power spectrum with the
Poisson self-pair contribution subtracted as
\be
\label{eqn:auto}
\begin{aligned}
P^{XX}(k,\mu)
={}&P^{XX,\mathrm{det}}(k,\mu;\,b_1^X,b_2^X,\ldots,c_0^X,c_2^X,c_4^X,\ctilde^X)\\
&+\frac{1}{\nbar_X}\left[\Pshot^X+
  \left(a_0^X+a_2^X\mu^2\right)\frac{k^2}{\knl^2}\right].
\end{aligned}
\ee
Here $k$ is the wavenumber, $\mu$ is the cosine of the wavevector's
angle to the line of sight, and $\nbar_X$ is the mean number density
of sample $X$.
The deterministic term contains the linear redshift-space spectrum
$(b_1^X+f\mu^2)^2P_{\mathrm{lin}}(k)$, the one-loop corrections, and
the counterterms. The bias coefficients $b_1,b_2$ describe the linear
and quadratic density responses, while $b_{\G_2},b_{\Gamma_3}$
multiply the quadratic and cubic tidal operators, respectively.
The counterterms $c_0,c_2,c_4$ absorb unresolved small-scale physics
through contributions proportional to $k^2P_{\mathrm{lin}}$, with different
angular dependences. We also retain the higher-order Fingers-of-God (FoG) term
proportional to $\ctilde k^4P_{\mathrm{lin}}$. Both are given explicitly in \cref{sec:eft_model}.
The stochastic terms describe departures from Poisson shot noise.
The subscripts on $a_0$ and $a_2$ label the $\mu^0$ and $\mu^2$ terms.
Their coefficients are dimensionless, with the fixed normalization
$\knl=0.7\,h_{\mathrm{ref}}\,\mathrm{Mpc}^{-1}$, where $h_{\mathrm{ref}}=0.6736$
is the dimensionless Hubble parameter of the fixed reference cosmology.
The subtracted Poisson term $1/\nbar_X$ is added once to the
diagonal field spectra used in the fiducial Gaussian covariance.}

The cross-spectrum between tracers $X$ and $Y$ is
\revchange{\be
\label{eqn:cross}
\begin{aligned}
P^{XY}(k,\mu)
={}&P^{XY,\mathrm{det}}(k,\mu;\,b_1^X,b_1^Y,\ldots,c_0^X,c_0^Y,\ldots,\ctilde^X,\ctilde^Y)\\
&+\frac{1}{\sqrt{\nbar_X\nbar_Y}}\left[\Pshot^{XY}
+a_2^{XY}\left(\frac{k\mu}{\knl}\right)^2\right].
\end{aligned}
\ee
The deterministic term uses the bias and counterterm coefficients of
both populations, shared with their auto-spectra.
The coefficients $\Pshot^{XY}$ and $a_2^{XY}$ are independent
of all auto-spectrum stochastic coefficients. We retain only the
constant and $(k\mu)^2$ terms of the stochastic expansion used by
Ebina and White~\cite{Ebina:2024ojt}. The $(k\mu)^2$ term describes
the leading correction from unresolved line-of-sight velocities.
We omit the independent isotropic $k^2$ term, motivated by the
$z=0$ mock tests of Mergulh\~ao et al.~\cite{Mergulhao:2023zso},
who found little change in cosmological constraints when including
all three cross-stochastic terms at $k_{\mathrm{max}}=0.14\iMpch$.
Disjoint catalogs have no shared-object Poisson noise, but residual
cross-stochasticity associated with halo occupation and exclusion
can remain~\cite{Mergulhao:2023zso}.
We set the fiducial values of $\Pshot^{XY}$ and $a_2^{XY}$
to zero and marginalize over both with Gaussian prior standard
deviations of one. The auto-spectrum stochastic coefficients have
the same fiducial values and prior widths.
This choice assumes that no galaxy appears in two catalogs, so that the
cross-spectra, including DESI-ELG\,$\times$\,PFS-ELG, carry no shared-object
Poisson noise at the fiducial point. \Cref{sec:results_sensitivity} relaxes
this assumption.

These expressions specify the intrinsic redshift-space spectra. The complete
modeled means, including the residual stochastic terms, undergo the
Alcock--Paczy\'nski (AP) mapping of \cref{sec:eft_model} before projection onto the observed
multipoles,
\[
P_\ell^{XY}(k)=\frac{2\ell+1}{2}
\int_{-1}^{1}\!\dd\mu\,L_\ell(\mu)P_{\mathrm{obs}}^{XY}(k,\mu).
\]
Here $L_\ell$ is the Legendre polynomial of order $\ell$.}

\revchange{Jointly fitting $P_\ell^{XX}$, $P_\ell^{YY}$ and
$P_\ell^{XY}$ provides different combinations of the same tracer
biases, together with the independent auto- and cross-stochastic
coefficients. The constant stochastic contribution is a monopole
offset at the fiducial geometry. The $k^2$ stochastic terms can be
correlated with the leading counterterms, but their shapes differ:
the latter contain $k^2P_{\mathrm{lin}}(k)$, while the former do not
multiply $P_{\mathrm{lin}}$. The nonlinear biases supply further
scale and angular dependence.}

\subsection{Joint Fisher across the partitioned footprint}
\label{sec:joint_fisher}

The joint multi-tracer Fisher operates across the full DESI
footprint with the volume partitioned by tracer availability. The
$14{,}000\;\mathrm{deg}^2$ DESI footprint splits into the
$1{,}200\;\mathrm{deg}^2$ \revchange{DESI\,$\times$\,PFS} overlap and the
$12{,}800\;\mathrm{deg}^2$ DESI-only non-overlap region. \revchange{For each nominal
redshift bin, the likelihood information is assembled as}
\be
\label{eq:volume_partition}
F_z = F_z^{\mathrm{overlap}}(V_{\mathrm{overlap}})
    + F_z^{\mathrm{nonoverlap}}(V_{\mathrm{nonoverlap}}),
\ee
\revchange{where $V_{\mathrm{overlap}}$ and $V_{\mathrm{nonoverlap}}$ are the comoving
volumes of these two regions, integrated over the populated radial intervals
in \cref{sec:surveys}.}
$F_z^{\mathrm{overlap}}$ includes \revchange{up to} 4 tracers (PFS-ELG plus
DESI-LRG, DESI-ELG, DESI-QSO when active in the bin) with their
auto- and cross-spectra\revchange{, while} $F_z^{\mathrm{nonoverlap}}$ includes only DESI tracers.
\revchange{PFS contributes below $z=1.6$ in the main comparison, and
\cref{sec:redshift} discusses its extension to $z=2.4$. The PFS-ELG
densities and bias follow the forecast of Takada et al.~\cite{Takada:2014pfs}
(\cref{sec:surveys}).}
\revchange{We approximate the angular and radial survey regions as
independent homogeneous volumes, neglecting survey-window mode coupling,
so their Fisher matrices add.
Within each region we use the Gaussian covariance of the included spectra and omit connected non-Gaussian terms,
including super-sample covariance (SSC). These omitted correlations
can change the marginalized errors and relative gain from cross-spectra
\cite{Wadekar:2019rdu}, and we have not evaluated them for our joint samples.
For BOSS, Wadekar et al.~\cite{Wadekar:2020hax} found that including
the connected trispectrum and SSC changed marginalized cosmological
errors by $\lesssim10\%$ at $k_{\mathrm{max}}=0.25\iMpch$.
Chudaykin et al.~\cite{Chudaykin:2025aux} further argued that comparable shot
noise and weaker nonlinear clustering at higher redshifts support the
Gaussian approximation for DESI DR1.
These results motivate the Gaussian approximation in our multi-tracer covariances and analyses at $\kmax=0.20\iMpch$.}

\revchange{We combine the per-$z$-bin Fishers across the eight nominal
redshift bins covering $z\in[0.4,2.1]$, with shared
cosmological parameters
$\boldsymbol{\theta}_{\mathrm{cos}}=
(\ln(10^{10}A_s),\,\Mnu,\,\Om)$ and per-tracer, per-bin nuisance parameters. Within each nominal bin, a
population shares its nuisance parameters between the overlap and
non-overlap regions, under the population-uniformity assumption of
\cref{sec:hod_uniformity}. In the main forecasts we fix
$h=0.6736$, $\omega_b=0.02237$ and $n_s=0.9649$.
The Gaussian prior standard deviations on $\Mnu$ and $\Om$
are $5\,\mathrm{eV}$ and $1$, respectively, with no
amplitude prior. These priors are added once to the combined
Fisher, together with the nuisance priors in
\cref{tab:eft_params}.

We report the derived growth predictions $G_a=f(z_a)\sigma_8(z_a)$
at the nominal-bin midpoints $z_a$. Their covariance follows from
the cosmological covariance $C_{\mathrm{cos}}^{\mathrm{marg}}$ after nuisance
marginalization,
\be
\label{eq:fsigma8_cov}
C^{G}_{ab}=\sum_{i,j}J_{ai}
 (C_{\mathrm{cos}}^{\mathrm{marg}})_{ij}J_{bj},
\qquad
J_{ai}=\frac{\partial[f(z_a)\sigma_8(z_a)]}
                 {\partial(\boldsymbol{\theta}_{\mathrm{cos}})_i}\,.
\ee
Here $a,b=1,\ldots,8$ label the nominal redshift bins, while $i,j=1,2,3$
label the entries of $\boldsymbol{\theta}_{\mathrm{cos}}$.
We quote
$\sigma[f\sigma_8(z=1.1)]$ for the headline comparison.}

\revchange{For DESI alone, we verify the volume partition by checking}
$F_z^{\mathrm{overlap}}+F_z^{\mathrm{nonoverlap}}=F_z(V_{\mathrm{overlap}}+V_{\mathrm{nonoverlap}})$ to numerical precision (see \cref{app:fisher_checks}).

\subsection{Analysis configurations}
\label{sec:scenarios}

We report \revchange{four cases at the common scale cut
$\kmax=0.20\iMpch$ (\cref{sec:asymmetric_kmax}), with identical redshift bins, survey
inputs and priors for the parameters shared by each comparison:
\begin{enumerate}
\item[(A)] \emph{DESI auto-spectra.} Each DESI
population contributes its auto-spectrum multipoles.
\item[(B)] \emph{DESI multi-tracer.} DESI cross-spectrum
multipoles are added to A.
\item[(C)] \emph{DESI multi-tracer plus PFS auto-spectra.}
PFS auto-spectrum multipoles are added to B in the overlap.
DESI\,$\times$\,PFS cross-spectra are not part of the observable vector.
\item[(D)] \emergencystretch=1em\relax\emph{DESI+PFS multi-tracer.} All available
DESI\,$\times$\,PFS cross-spectrum multipoles are added to C.
\end{enumerate}
Each retained spectrum contributes $\ell=0,2,4$.
All four cases use the full Gaussian covariance, including contributions
from cross-spectra absent from the observable vector, as in A and C. Each included cross pair introduces the two
stochastic nuisance parameters defined in
\cref{sec:cross_power}.

Comparing A with B isolates the
addition of DESI cross-spectra. Comparing B with C isolates
the added PFS auto-spectra, and C with D the added
DESI\,$\times$\,PFS cross-spectra, including the cost of their
additional stochastic parameters. The B--D comparison gives the total incremental PFS contribution.
In A, each tracer retains its own fiducial bias and
nuisance parameters. No density-weighted effective bias is
needed because no merged catalog is formed.}

\subsection{Comparison with simulation-based priors}
\label{sec:sbp_comparison}

\revchange{SBPs encode the joint distribution of EFT parameters inferred
from HOD mocks, through power-spectrum fits as in
Zhang et al.~\cite{Zhang:2024thl} or field-level fits as in
Chudaykin et al.~\cite{Chudaykin:2026nls}. Their} accuracy is bounded
by the fidelity of the galaxy--halo connection model.
\revchange{The joint multi-tracer likelihood constrains the same parameters
directly from the auto- and cross-spectra.
We keep the input nuisance priors fixed and do not reuse these likelihood
constraints as independent priors on the same data.
For the parameters that the likelihood constrains well, the two approaches
can check each other when sample selection, parameter definitions and
modeling assumptions are matched (\cref{sec:discussion}).}

\section{Fisher forecast setup}
\label{sec:fisher}

\revchange{We forecast cosmological constraints for DESI and its overlap with PFS
using the joint Fisher formalism above. Below we specify the survey inputs,
the one-loop model and its AP mapping, the nuisance parameters,
priors and fiducials, the scale cuts, the numerical Fisher calculation and the
population-uniformity assumption.}

\subsection{Survey specifications}
\label{sec:surveys}

\paragraph{DESI.}\leavevmode%
\revchange{We adopt a $14{,}000\;\mathrm{deg}^2$ DESI forecast
footprint~\cite{DESI:2016fyo}, with three tracer populations.}
The LRG sample covers $z\in[0.4,1.1]$ with
$b_1(z)=1.7/D(z)$\revchange{, the highest bias of the three.
The QSO sample spans $z\in[0.8,2.1]$ with} $b_1(z)=1.2/D(z)$\revchange{,
and the ELG sample spans $z\in[0.8,1.6]$ with} $b_1(z)=0.84/D(z)$\revchange{,
the lowest bias of the three.
Here we normalize the growth factor such that $D(z=0.0)=1$.
The DESI forecast number-density
tables have native widths $\Delta z=0.01$ for LRGs and ELGs,
and $0.02$ for QSOs.}

\paragraph{PFS.}\leavevmode%
\revchange{We place a $1{,}200\;\mathrm{deg}^2$ PFS-ELG footprint inside
the DESI footprint and adopt $b_1^{\mathrm{PFS}}(z)=0.9+0.4z$,
the fiducial relation used in the PFS
forecast of Takada et al.~\cite{Takada:2014pfs}. We adopt the
number densities in their Table~2, taken to be constant within
its seven intervals over $z\in[0.6,2.4]$.
In the main comparison we retain
PFS below $z=1.6$, where it overlaps the DESI galaxy samples. Above this
redshift, only the sparse DESI-QSO sample overlaps PFS, so additional PFS
information enters mainly through its auto-spectra. The separate high-redshift
study extends PFS to $z=2.1$ alongside DESI-QSO, and then to $z=2.4$ with a
PFS-only tail. \Cref{sec:redshift} quantifies the additional constraints from
this extension of PFS redshift coverage.
PFS-ELG and DESI-ELG have separate nuisance parameters.}

\paragraph{Overlap geometry.}\leavevmode%
\revchange{The eight nominal redshift bins are
\[
\begin{gathered}\relax
[0.4,0.6],\quad [0.6,0.8],\quad [0.8,1.0],\quad [1.0,1.2],\\
[1.2,1.4],\quad [1.4,1.6],\quad [1.6,2.0],\quad [2.0,2.1].
\end{gathered}
\]
The five bins from $0.6$ to $1.6$ follow the bin edges
of the PFS forecast of Takada et al.~\cite{Takada:2014pfs}.
The nominal bins define the nuisance-parameter sets and the
model redshifts, taken to be their midpoints. All cases in \cref{sec:scenarios} use this same grid for their common
populations. We label the subsamples LRG-1--LRG-4, ELG-1--ELG-4,
QSO-1--QSO-6 and PFS-ELG-1--PFS-ELG-5, numbering each population
by increasing nominal redshift. These are forecast-specific labels
and do not in general match the sample definitions or bin boundaries
of DESI DR1 or DR2. The high-redshift extension adds PFS-ELG-6,
PFS-ELG-7 and PFS-ELG-8 over $[1.6,2.0]$, $[2.0,2.1]$ and
$[2.1,2.4]$, respectively.

For LRG-2, we retain separate volume-weighted densities below
and above $z=0.7$, with PFS-ELG-1 present throughout.
We also split $[1.0,1.2]$ at the LRG-4 upper edge, $z=1.1$.
Only the populated intervals contribute
to each auto- or cross-spectrum. Within a nominal bin, these
radial intervals retain the same subsample label, model redshift
and nuisance parameters for each population. Each density is a volume-weighted mean
of the native table over the populated interval. Comoving
volumes are integrated over that interval and scaled to the
overlap or non-overlap area. \Cref{tab:surveys} lists the
intervals used in the main overlap comparison.

For the tracers considered here, the official DESI DR1
full-shape analysis uses three LRG bins, $[0.4,0.6]$,
$[0.6,0.8]$ and $[0.8,1.1]$, one ELG bin
$[1.1,1.6]$, and one QSO bin
$[0.8,2.1]$~\cite{DESI:2024hhd}. Its catalog definitions
also include DESI DR1 ELG1 over $[0.8,1.1]$, but this sample was
excluded from the full-shape fit because of residual
systematics~\cite{DESI:2024jxi}. Our main forecast includes
this interval. The matched DESI and DESI+PFS sensitivity
comparison removes it. The chosen subdivision resolves the changing survey overlap. Our comparisons
use the same grid and assume that residual selection effects within each bin
have been corrected. We have not quantified the gain from changing the
tomographic grid.}

\begin{table}[t]
\centering
\small
\setlength{\tabcolsep}{3pt}
\begin{tabular}{llrrrrrrrrrrr}
\hline
 & & & \multicolumn{2}{c}{PFS-ELG} & \multicolumn{2}{c}{DESI-ELG} & \multicolumn{2}{c}{DESI-LRG} & \multicolumn{2}{c}{DESI-QSO} & & \\
\revchange{PFS sample} & \revchange{$z$ interval} & \revchange{$z_{\mathrm{model}}$} & $\nbar$ & $b_1$ & $\nbar$ & $b_1$ & $\nbar$ & $b_1$ & $\nbar$ & $b_1$ & $N_t$ & $N_{\mathrm{spec}}$ \\
\hline
\revchange{PFS-ELG-1} & \revchange{$[0.6,\,0.7]$} & \revchange{0.7} & \revchange{1.90} & \revchange{1.18} & \revchange{---} & \revchange{---} & \revchange{5.68} & \revchange{2.44} & \revchange{---} & \revchange{---} & \revchange{2} & \revchange{3} \\
\revchange{PFS-ELG-1} & \revchange{$[0.7,\,0.8]$} & \revchange{0.7} & \revchange{1.90} & \revchange{1.18} & \revchange{---} & \revchange{---} & \revchange{4.69} & \revchange{2.44} & \revchange{---} & \revchange{---} & \revchange{2} & \revchange{3} \\
\revchange{PFS-ELG-2} & \revchange{$[0.8,\,1.0]$} & \revchange{0.9} & \revchange{6.00} & \revchange{1.26} & \revchange{8.14} & \revchange{1.32} & \revchange{2.26} & \revchange{2.68} & \revchange{0.29} & \revchange{1.89} & \revchange{4} & \revchange{10} \\
\revchange{PFS-ELG-3} & \revchange{$[1.0,\,1.1]$} & \revchange{1.1} & \revchange{5.80} & \revchange{1.34} & \revchange{5.00} & \revchange{1.45} & \revchange{0.39} & \revchange{2.93} & \revchange{0.29} & \revchange{2.07} & \revchange{4} & \revchange{10} \\
\revchange{PFS-ELG-3} & \revchange{$[1.1,\,1.2]$} & \revchange{1.1} & \revchange{5.80} & \revchange{1.34} & \revchange{4.20} & \revchange{1.45} & \revchange{---} & \revchange{---} & \revchange{0.30} & \revchange{2.07} & \revchange{3} & \revchange{6} \\
\revchange{PFS-ELG-4} & \revchange{$[1.2,\,1.4]$} & \revchange{1.3} & \revchange{7.80} & \revchange{1.42} & \revchange{2.78} & \revchange{1.57} & \revchange{---} & \revchange{---} & \revchange{0.32} & \revchange{2.24} & \revchange{3} & \revchange{6} \\
\revchange{PFS-ELG-5} & \revchange{$[1.4,\,1.6]$} & \revchange{1.5} & \revchange{5.50} & \revchange{1.50} & \revchange{1.24} & \revchange{1.69} & \revchange{---} & \revchange{---} & \revchange{0.31} & \revchange{2.42} & \revchange{3} & \revchange{6} \\
\hline
\end{tabular}
\caption{\revchange{Survey inputs in the $1{,}200\;\mathrm{deg}^2$
overlap footprint for the main comparison. Densities are} in
units of $10^{-4}\,h^3\,\mathrm{Mpc}^{-3}$ \revchange{and are
volume-weighted over the listed radial intervals.
The model redshift is the midpoint of the parent nominal bin.} $N_t$ is the number of overlapping tracers\revchange{, and}
$N_{\mathrm{spec}}=N_t(N_t+1)/2$ is the number of auto- and
cross-spectra \revchange{in case D.}
The LRG sample drops out above $z=1.1$. \revchange{The DESI densities and
biases are also used in the non-overlap region. Volumes use
the fixed reference cosmology.}}
\label{tab:surveys}
\end{table}

\subsection{\texorpdfstring{\revchange{One-loop model and AP mapping}}{One-loop model and AP mapping}}
\label{sec:eft_model}

{\emergencystretch=3em
\revchange{We compute the one-loop redshift-space auto- and cross-spectra
with the Eulerian perturbation-theory (EPT) calculation in
\texttt{ELF}\footnote{\url{https://github.com/archaeo-pteryx/elf}},
the differentiable JAX implementation of Eulerian and Lagrangian
perturbation theory by Kobayashi and Akitsu~\cite{Kobayashi:2026eee}.
We use its FFTLog-based Hankel method to evaluate the loop integrals.
The linear total-matter spectrum and $\sigma_8(z)$ are emulated
using} \texttt{cosmopower-jax}\footnote{\url{https://github.com/dpiras/cosmopower-jax}}, the JAX-based \texttt{CosmoPower} interface in Piras and Spurio Mancini~\cite{Piras:2023aub}
\revchange{with the $\nu\Lambda$CDM} networks\footnote{\url{https://github.com/cosmopower-organization/jense_2024_emulators/tree/main/jense_2023_camb_mnu/networks}} of Jense
et al.~\cite{Jense:2024llt}.
\revchange{The calculation
does not include IR resummation or neutrino-specific loop
dynamics.
The growth rate $f=\dd\ln D/\dd\ln a$, where $D$ is the
linear growth factor and $a$ the scale factor, is evaluated
from the scale-independent Heath-integral solution for a flat
matter--$\Lambda$ background, with neutrino mass entering only
through the background density.}\par}

\revchange{Following Mergulh\~ao et al.~\cite{Mergulhao:2023zso}
and Rubira and Conteddu~\cite{Rubira:2025scu}, we assign the leading
counterterms to the tracer density fields,
$\delta_X\supset k^2D_X\delta_{\mathrm{lin}}$, where $\delta_{\mathrm{lin}}$ is the linear
matter density and $D_X=d_0^X+d_2^X\mu^2+d_4^X\mu^4$. Writing $Z_X=b_1^X+f\mu^2$,
their contributions to the auto- and cross-spectra are
\[
P_{\mathrm{ctr}}^{XY}(k,\mu)=k^2P_{\mathrm{lin}}(k)
\left(Z_XD_Y+Z_YD_X\right).
\]
At each cosmology and nuisance point, the three $d_i^X$ are fixed by
the intrinsic auto-spectrum multipoles,
\[
\left(P_{0,\mathrm{ctr}}^{XX},P_{2,\mathrm{ctr}}^{XX},P_{4,\mathrm{ctr}}^{XX}\right)
=-2k^2P_{\mathrm{lin}}(k)
\left(c_0^X,\frac{2f}{3}c_2^X,\frac{8f^2}{35}c_4^X\right).
\]
The subscripts on the counterterms $c_\ell^X$ denote their Legendre orders,
while those on $d_i^X$ denote powers of $\mu$.
Elsewhere we suppress the tracer label $X$ on $c_\ell^X$.
As in Rubira and Conteddu, $D_X$ has no $\mu^6$ term, but its $\mu^4$ term
gives $Z_XD_Y$ a $\mu^6$ part. Each spectrum therefore also receives an
$\ell=6$ contribution, which we retain in the AP mapping and the covariance.
For the higher-order FoG term, we use the field contribution
$\delta_X\supset-\ctilde^Xk^4f^4\mu^4Z_X\delta_{\mathrm{lin}}/2$~\cite{Ivanov:2024xgb},
giving
\[
P_{\mathrm{FoG}}^{XY}(k,\mu)=-\frac{\ctilde^X+\ctilde^Y}{2}\,
k^4f^4\mu^4Z_XZ_YP_{\mathrm{lin}}(k).
\]}

\revchange{For the AP mapping in \cref{sec:cross_power}, we write
$\alpha_\perp=D_A(z)/D_A^{\mathrm{ref}}(z)$ and
$\alpha_\parallel=H^{\mathrm{ref}}(z)/H(z)$, so that
\[
k'=k\left[\frac{1-\mu^2}{\alpha_\perp^2}
          +\frac{\mu^2}{\alpha_\parallel^2}\right]^{1/2},
\qquad
\mu'=\frac{k\mu}{\alpha_\parallel k'},
\qquad
P_{\mathrm{obs}}^{XY}(k,\mu)=
\frac{P_{\mathrm{true}}^{XY}(k',\mu')}
     {\alpha_\perp^2\alpha_\parallel}\,.
\]
Here $D_A$ is the angular-diameter distance and $H$ the Hubble rate.
The labels $\perp$ and $\parallel$ refer to directions transverse
and parallel to the line of sight. The unprimed $(k,\mu)$ are defined
in the reference cosmology, while $(k',\mu')$ are their values in the
trial cosmology.
Observed wavenumbers, number densities and survey volumes
are held in the reference cosmology, with fixed $h_{\mathrm{ref}}=0.6736$
units, so $h$ in all wavenumber, length and density units denotes $h_{\mathrm{ref}}$. In unweighted catalogs, objects shared by two samples, with number density
$\nbar_{X\cap Y}$, add $\nbar_{X\cap Y}/\sqrt{\nbar_X\nbar_Y}$ to the constant term of
\cref{eqn:cross}, which we include only in the shared-ELG variant of
\cref{sec:results_sensitivity}. At fixed observed counts this term has no AP
response: the true densities scale as $1/J$, with $J=\alpha_\perp^2\alpha_\parallel$, so the
true Poisson power $\nbar_{X\cap Y}/(\nbar_X\nbar_Y)$ scales as $J$ and cancels the
factor $1/J$ above.}

\subsection{EFT parameterization}
\label{sec:eft_params}

\revchange{We use amplitude-scaled nuisance coordinates, as in
Chudaykin et al.~\cite{Chudaykin:2025aux}.
Define
\[
\mathcal A(z)\equiv\left[\frac{\sigma_8(z)}{\sigma_{8,\mathrm{ref}}(z)}\right]^2,
\]
where the denominator is evaluated in the fixed reference cosmology.
The nuisance coordinates held fixed when varying cosmology are
$b_1\se$, $b_2\se^2$, $b_{\G_2}\se^2$,
$\mathcal A b_{\Gamma_3}$, $\mathcal A c_0$,
$\mathcal A c_2$, $\mathcal A c_4$,
$\mathcal A^2\ctilde$, and the three dimensionless stochastic
coefficients. A cosmological variation therefore changes the
corresponding unscaled coefficients even when these nuisance
coordinates are fixed.
The counterterm field coefficients $d_i^X$ of \cref{sec:eft_model} also depend
on $b_1^X$ and $f$, so they change with cosmology at fixed nuisance coordinates.
All active nuisance parameters enter the Fisher matrix explicitly,
with the priors in \cref{tab:eft_params}.
We impose no informative prior on $b_1\se$.

Each tracer has an independent nuisance set in every nominal
redshift bin. Within one bin, the same population shares its
parameters across overlap and non-overlap regions and any radial
subdivision used to account for its redshift coverage. Each included cross-pair has its own two stochastic
parameters per nominal bin. The DESI multi-tracer forecast (B) has
14 tracer--bin sets and 8 cross-pair--bin sets, giving
$3+14\times11+8\times2=173$ parameters. The full DESI+PFS forecast (D), with
PFS over its populated support $0.6\leq z\leq1.6$, has 19 tracer--bin sets
and 19 cross-pair--bin sets, or $3+19\times11+19\times2=250$
parameters. After combining the independent survey regions and radial
cells, each distinct nuisance parameter receives its prior once.}

\begin{table}[t]
\centering
\small
\begin{tabular}{lrrrrrl}
\hline
Parameter & \revchange{PFS-ELG-2} & \revchange{ELG-1} & \revchange{LRG-3} & \revchange{QSO-1} & \revchange{Prior s.d.} & Units \\
\hline
$b_1\se$ & 0.65 & 0.68 & 1.38 & 0.97 & \revchange{flat} & --- \\
$b_2\se^2$ & $-0.21$ & $-0.21$ & 0.40 & $-0.07$ & \revchange{5} & --- \\
$b_{\G_2}\se^2$ & $-0.02$ & $-0.02$ & $-0.13$ & $-0.07$ & \revchange{5} & --- \\
\revchange{$\mathcal A b_{\Gamma_3}$} & \revchange{0.14} & \revchange{0.18} & \revchange{0.92} & \revchange{0.49} & \revchange{1} & \revchange{---} \\
\revchange{$\mathcal A c_0$} & \revchange{0} & \revchange{0} & \revchange{0} & \revchange{0} & \revchange{30} & \revchange{$[h_{\mathrm{ref}}^{-1}\mathrm{Mpc}]^2$} \\
\revchange{$\mathcal A c_2$} & \revchange{29} & \revchange{30} & \revchange{62} & \revchange{44} & \revchange{30} & \revchange{$[h_{\mathrm{ref}}^{-1}\mathrm{Mpc}]^2$} \\
\revchange{$\mathcal A c_4$} & \revchange{0} & \revchange{0} & \revchange{0} & \revchange{0} & \revchange{30} & \revchange{$[h_{\mathrm{ref}}^{-1}\mathrm{Mpc}]^2$} \\
\revchange{$\mathcal A^2\ctilde$} & \revchange{127} & \revchange{400} & \revchange{400} & \revchange{400} & \revchange{400} & \revchange{$[h_{\mathrm{ref}}^{-1}\mathrm{Mpc}]^4$} \\
$\Pshot$ & 0 & 0 & 0 & 0 & \revchange{1} & --- \\
$a_0$ & 0 & 0 & 0 & 0 & \revchange{1} & --- \\
$a_2$ & 0 & 0 & 0 & 0 & \revchange{1} & --- \\
\hline
\end{tabular}
\caption{\color{RevisionColor}The 11 nuisance coordinates per tracer and redshift bin,
with fiducial values at $z=0.9$ and baseline Gaussian prior standard
deviations. At the reference cosmology, $\mathcal A=1$.
The fiducial prescriptions are given in \cref{sec:eft_fiducials}.
The prior widths apply independently to each tracer--bin set.
The flat prior on $b_1\se$ contributes no information to the Fisher matrix.
Each included cross-pair adds two dimensionless stochastic
parameters, $\Pshot^{XY}$ and $a_2^{XY}$, both with fiducial
value zero and prior standard deviation one.}
\label{tab:eft_params}
\end{table}

\subsection{\texorpdfstring{\revchange{EFT fiducials}}{EFT fiducials}}
\label{sec:eft_fiducials}

\revchange{The linear bias follows the survey
model in \cref{sec:surveys}. For all populations, the nonlinear-bias
fiducials are
\[
\begin{gathered}
b_2=0.412-2.143b_1+0.929b_1^2+0.008b_1^3,\\
b_{\G_2}=-\frac{2}{7}(b_1-1),\qquad
b_{\Gamma_3}=\frac{23}{42}(b_1-1).
\end{gathered}
\]
The $b_2$ relation is the fitting formula of
Lazeyras et al.~\cite{Lazeyras:2015lgp}. The tidal-bias relations follow
from local Lagrangian co-evolution, as reviewed by
Desjacques et al.~\cite{Desjacques:2016bnm}. The bias parameters are
varied independently in the Fisher analysis.

At the reference cosmology, we set
$\ctilde=400\,[h_{\mathrm{ref}}^{-1}\mathrm{Mpc}]^4$ for all DESI
populations. This value is the benchmark of
Chudaykin et al.~\cite{Chudaykin:2025aux}, adopted here as an
assumption for each tracer. For PFS we use
$\ctilde^{\PFS}=400\,\rsv^4\,[h_{\mathrm{ref}}^{-1}\mathrm{Mpc}]^4$
with $\rsv=0.75$, giving $126.5625\,[h_{\mathrm{ref}}^{-1}\mathrm{Mpc}]^4$.
The assumed ratio $\rsv\equiv\sigma_{v,\PFS}/\sigma_{v,\mathrm{DESI\text{-}ELG}}$
sets this heuristic velocity-dispersion scaling.

We set $c_0=c_4=0$ and use
$c_2^{\mathrm{DESI\text{-}ELG}}=30\,[h_{\mathrm{ref}}^{-1}\mathrm{Mpc}]^2$.
For DESI-LRG and DESI-QSO we take
$c_2^X=30(b_1^X/1.3)\,[h_{\mathrm{ref}}^{-1}\mathrm{Mpc}]^2$.
For PFS we use
$c_2^{\PFS}=30(b_1^{\PFS}/b_{1,\mathrm{ref}})\,[h_{\mathrm{ref}}^{-1}\mathrm{Mpc}]^2$,
where $b_{1,\mathrm{ref}}$ is the fiducial DESI-ELG bias when that
population is present in the bin, and 1.3 otherwise. The
counterterm combinations in \cref{tab:eft_params} are varied independently.}

\subsection{\texorpdfstring{\revchange{Scale cuts}}{Scale cuts}}
\label{sec:asymmetric_kmax}

\revchange{Auto- and cross-spectra use the same observed wavenumber range
in every redshift bin,
\be
\label{eqn:kmax_asym}
0.01\iMpch \leq k \leq \kmax=0.20\iMpch\,.
\ee
The DESI DR1 full-shape analysis~\cite{DESI:2024jxi} uses the same upper cut
for the monopole and quadrupole, whereas we also retain the hexadecapole. One-loop accuracy depends on
redshift, tracer bias and velocity dispersion, so the smaller
assumed PFS FoG coefficient alone does not justify extending
its $k$-range. We assess scale sensitivity by repeating the DESI
and DESI+PFS multi-tracer analyses with $\kmax=0.15\iMpch$.}

\subsection{Multi-tracer Fisher matrix}
\label{sec:mt_fisher}

The observables are power spectrum multipoles
$P_\ell(k)$ for $\ell=0,2,4$~\cite{Ivanov:2019pdj,DAmico:2019fhj},
computed for each auto- and cross-spectrum pair. At each wavenumber $k$, the observable vector has $3 N_{\mathrm{spec}}$ elements, where $N_{\mathrm{spec}}=N_t(N_t{+}1)/2$ is the number of auto- and cross-spectrum pairs \revchange{for $N_t$ active populations}, up to 30 elements for 4 tracers.

The Gaussian covariance at each $k$ is
\be
\Cov\!\left[P_\ell^{XY},\, P_{\ell'}^{WZ}\right]\!(k)
= \frac{(2\ell{+}1)(2\ell'{+}1)}{2 N_{\mathrm{modes}}(k)}
  \int_{-1}^{1}\!\dd\mu\; L_\ell(\mu)\, L_{\ell'}(\mu)\,
  \bigl[P^{XW} P^{YZ}
  + P^{XZ} P^{YW}\bigr],
\ee
{\emergencystretch=1em
\revchange{where $X,Y,W,Z$ label tracer populations and $\ell,\ell'=0,2,4$
label the multipole orders. The mode count}
$N_{\mathrm{modes}}(k)=k^2\Delta k\,V/(2\pi^2)$ \revchange{refers to a shell
of width $\Delta k$ in volume $V$. Here $P^{XY}$ is the total fiducial
spectrum: the one-loop model, including its stochastic terms,
plus $\delta_{XY}^{\mathrm{K}}/\nbar_X$, where $\delta_{XY}^{\mathrm{K}}$ is the Kronecker delta.
The fiducial cross-stochastic
coefficients are zero, while their derivatives remain in the
Fisher matrix (\cref{sec:cross_power}). The factor $1/2$
corresponds to the integral over $-1\leq\mu\leq1$. In the
isotropic single-tracer limit this gives
$\operatorname{Var}P_0=2(P+1/\nbar)^2/N_{\mathrm{modes}}$.
We hold the covariance at the fiducial model and use
\[
F_{ij}=\sum_{k\ {\mathrm{shells}}}
\left(\frac{\partial\boldsymbol{P}}{\partial\theta_i}\right)^{\!\mathrm{T}}
 C^{-1}(k)
\left(\frac{\partial\boldsymbol{P}}{\partial\theta_j}\right).
\]
Here $\boldsymbol{P}(k)$ contains the retained spectrum multipoles,
$C(k)$ is their covariance, and $i,j$ run over all varied cosmological
and nuisance parameters $\boldsymbol{\theta}$.
The shell width is already included through $N_{\mathrm{modes}}$.

The angular integrals use 64-point Gauss--Legendre quadrature.
We use the same mean model in the
covariance and its parameter derivatives.
The Poisson term in the covariance
uses the fixed observed number density.
We compute derivatives of the multipoles with JAX automatic
differentiation. A cosmological variation changes the linear
spectrum, $\sigma_8(z)$, the growth rate, the loop terms and
the AP distances. The nuisance
coordinates in \cref{tab:eft_params} are held fixed in taking
these derivatives. We check these
derivatives against finite differences (\cref{app:derivatives}) and assess numerical convergence with respect to the loop, angular and wavenumber
grids.\par}}

\subsection{\texorpdfstring{\revchange{Population-uniformity assumption}}{Population-uniformity assumption}}
\label{sec:hod_uniformity}

\revchange{The volume-partitioned Fisher analysis shares the bias, counterterm
and normalized stochastic coefficients of a given DESI
population and nominal redshift bin between the overlap and
non-overlap regions.}
DESI targets are drawn from a
heterogeneous combination of input imaging
surveys with distinct photometric
systems~\cite{DESI:2022gle,Raichoor:2022jab,Chaussidon:2022pqg},
\revchange{and the DESI\,$\times$\,PFS} overlap covers only part of the full footprint\revchange{.
The two regions can sample populations with different clustering statistics. For a uniformly selected population at fixed
redshift, statistical homogeneity motivates common ensemble clustering
statistics. We assume that selection corrections recover this condition.
Sharing the coefficients then allows DESI observations outside the overlap to
constrain the same nuisance parameters that enter the overlap spectra.
This assumption does not impose an HOD model or an HOD-derived prior.
Different populations, including PFS-ELG and DESI-ELG, and
different nominal redshift bins retain independent
nuisance parameters.}

\section{Results}
\label{sec:results}

\revchange{Using the cases in \cref{sec:scenarios}, we examine how
the nuisance and cosmological constraints change with the treatment
of DESI correlations and the addition of PFS. We then assess the
remaining nuisance-marginalization cost and the sensitivity of the
gains to the analysis choices.}

\subsection{Per-redshift nuisance constraints}
\label{sec:results_priors}

\begin{figure}[t]
\centering
\includegraphics[width=\textwidth]{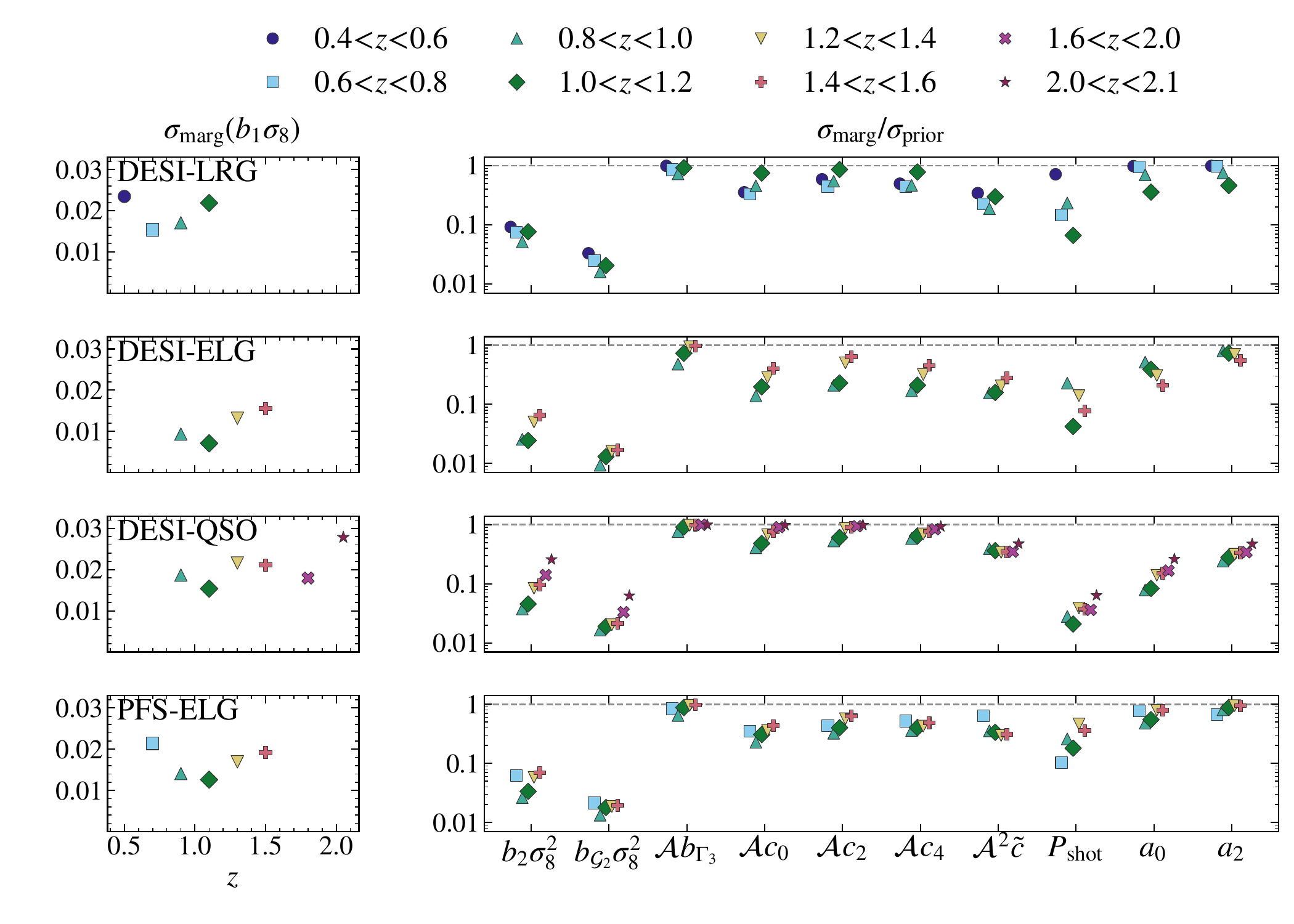}
\caption{\color{RevisionColor}Marginalized nuisance constraints from the full DESI+PFS
auto- and cross-spectrum analysis (case D), with
$k_{\mathrm{max}}=0.20\iMpch$ and the priors in
\cref{tab:eft_params}. Rows identify populations and symbols
identify nominal redshift bins. Their populated subintervals
are specified in \cref{sec:surveys}. Left: absolute
$1\sigma$ uncertainties on $b_1\se$, with a flat prior.
Right: posterior-to-prior width
ratios for the ten remaining per-population coordinates.
The dashed line marks unity. The amplitude factors
$\mathcal A$ follow \cref{sec:eft_params}.
Every point is marginalized over the other parameters in
the full survey fit, including all cross-stochastic
coefficients, which are not plotted.}
\label{fig:calibrated_vs_broad}
\end{figure}

\revchange{To show which nuisance parameters are constrained by the joint
clustering analysis, \Cref{fig:calibrated_vs_broad} compares their
marginalized uncertainties in case D with the prior
widths in \cref{tab:eft_params}. Each occupied population and
nominal redshift bin is shown separately: LRG-1--LRG-4,
ELG-1--ELG-4, QSO-1--QSO-6 and PFS-ELG-1--PFS-ELG-5. The errors
come from the full survey fit, with the three cosmological
parameters and all other nuisance parameters marginalized.

The linear-bias coordinate $b_1\se$ has
$\sigma(b_1\se)\simeq 0.007\text{--}0.028$ across the
displayed populations and bins. For the remaining coordinates,
we show $\sigma_{\mathrm{marg}}/\sigma_{\mathrm{prior}}$. The nonlinear biases $b_2\se^2$ and
$b_{\G_2}\se^2$ are substantially tightened in this joint
fit. For the latter, the ratio ranges from 0.009 to 0.063.
The counterterms and stochastic coefficients show a stronger
dependence on population and redshift. For example, the ratio
for $a_2$ is $0.24\text{--}0.47$ for QSOs and
$0.67\text{--}0.95$ for PFS-ELGs, while
$\mathcal A b_{\Gamma_3}$ remains close to its prior width
in several bins.

LRG-1 at $z\in[0.4,0.6]$ contributes only an auto-spectrum,
but its displayed uncertainty also benefits from the cosmological
constraints supplied by other bins. }

\subsection{Cosmology constraints}
\label{sec:results_full_area}

\begin{figure}[t]
\centering
\includegraphics[width=\textwidth]{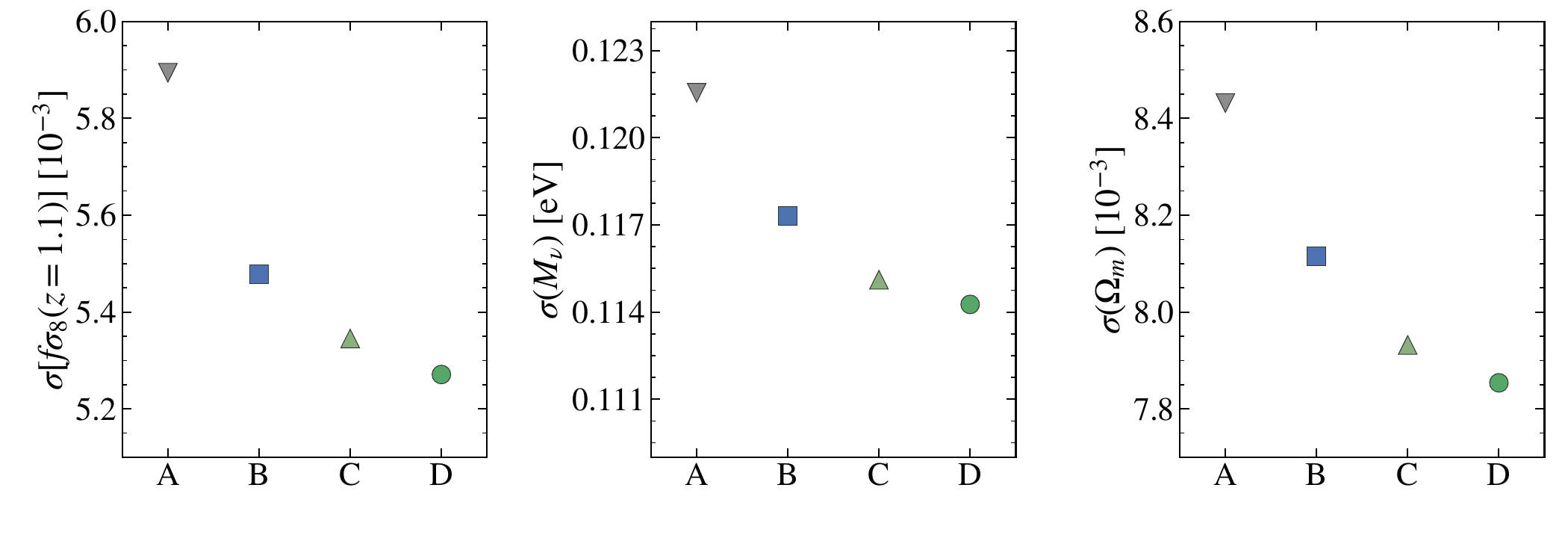}
\caption{\color{RevisionColor}Marginalized cosmological uncertainties for the matched
A/B/C/D forecasts at $\kmax=0.20\iMpch$.
The panels show the derived $f\se(z=1.1)$, neutrino mass
$\Mnu$, and matter density $\Om$.
A uses DESI auto-spectra.
B adds DESI cross-spectra, C the PFS auto-spectra, and D the DESI\,$\times$\,PFS cross-spectra.
All cases use the same eight nominal bins and common DESI
sample properties over $14{,}000\;\mathrm{deg}^2$.
PFS occupies $1{,}200\;\mathrm{deg}^2$ within this footprint,
with baseline support $0.6\leq z\leq1.6$.
All nuisance parameters of each case are marginalized with
the priors in \cref{tab:eft_params}, and $h,\omega_b,n_s$ are fixed.
Values and successive fractional changes are given in
\cref{tab:headline}.}
\label{fig:full_area}
\end{figure}

\revchange{\Cref{fig:full_area} and \cref{tab:headline} present the
cosmology forecasts for the four cases defined in \cref{sec:scenarios},
at $\kmax=0.20\iMpch$. The reported errors marginalize over
all nuisance parameters and the other varied cosmological parameters,
with $h,\omega_b,n_s$ fixed. Growth denotes the derived
$f\se(z=1.1)$ as defined in \cref{sec:joint_fisher}.
\Cref{tab:dr2_samples} lists the DESI sample properties on the
common grid, and \cref{tab:surveys} gives the PFS inputs.}

\begin{table}[t]
\centering\color{RevisionColor}
\small
\setlength{\tabcolsep}{4pt}
\begin{tabular}{lccccrr}
\hline
Sample & Nominal $z$ & Populated $z$ & $z_{\mathrm{model}}$ & $\nbar$ & $b_1$ & $V$ \\
 & interval & interval & & $[10^{-4}h^3\,\mathrm{Mpc}^{-3}]$ & & $[(h^{-1}\mathrm{Gpc})^3]$ \\
\hline
LRG-1 & $[0.4,0.6]$ & $[0.4,0.6]$ & 0.50 & 5.514 & 2.211 & 3.364 \\
LRG-2 & $[0.6,0.8]$ & $[0.6,0.8]$ & 0.70 & 5.141 & 2.441 & 5.220 \\
LRG-3 & $[0.8,1.0]$ & $[0.8,1.0]$ & 0.90 & 2.263 & 2.680 & 6.856 \\
LRG-4 & $[1.0,1.2]$ & $[1.0,1.1]$ & 1.10 & 0.393 & 2.926 & 3.946 \\
ELG-1 & $[0.8,1.0]$ & $[0.8,1.0]$ & 0.90 & 8.143 & 1.324 & 6.856 \\
ELG-2 & $[1.0,1.2]$ & $[1.0,1.2]$ & 1.10 & 4.585 & 1.446 & 8.184 \\
ELG-3 & $[1.2,1.4]$ & $[1.2,1.4]$ & 1.30 & 2.780 & 1.569 & 9.205 \\
ELG-4 & $[1.4,1.6]$ & $[1.4,1.6]$ & 1.50 & 1.238 & 1.695 & 9.954 \\
QSO-1 & $[0.8,1.0]$ & $[0.8,1.0]$ & 0.90 & 0.293 & 1.892 & 6.856 \\
QSO-2 & $[1.0,1.2]$ & $[1.0,1.2]$ & 1.10 & 0.296 & 2.065 & 8.184 \\
QSO-3 & $[1.2,1.4]$ & $[1.2,1.4]$ & 1.30 & 0.320 & 2.242 & 9.205 \\
QSO-4 & $[1.4,1.6]$ & $[1.4,1.6]$ & 1.50 & 0.308 & 2.421 & 9.954 \\
QSO-5 & $[1.6,2.0]$ & $[1.6,2.0]$ & 1.80 & 0.287 & 2.694 & 21.302 \\
QSO-6 & $[2.0,2.1]$ & $[2.0,2.1]$ & 2.05 & 0.260 & 2.923 & 5.495 \\
\hline
\end{tabular}
\caption{\color{RevisionColor}DESI sample properties on the nominal grid used
in all four forecasts. Each row has an independent nuisance
vector. The spectra and fiducial biases are evaluated at the
nominal-bin midpoint $z_{\mathrm{model}}$, including the partly
populated LRG-4 bin $[1.0,1.2]$, whose actual support ends at
$z=1.1$. The listed $\nbar$ is the volume-weighted mean of
the radial-interval densities in that row. The Fisher
calculation retains the separate intervals at $z=0.7$ and
$z=1.1$ described in \cref{tab:surveys}, with shared nuisance
parameters within the same population and nominal bin.
The volume $V$ covers the populated interval over the full
$14{,}000\;\mathrm{deg}^2$ footprint at the fiducial cosmology.
Overlapping sample volumes cannot simply be added.}
\label{tab:dr2_samples}
\end{table}
\revchange{Adding DESI cross-spectra (A to B) reduces the errors on
$f\se(z=1.1),\Mnu,\Om$ by $7.1\%,3.5\%,3.8\%$.
B includes the cross-spectra of the DESI samples in \cref{tab:dr2_samples} that overlap in each nominal redshift bin.

Adding PFS auto-spectra (B to C) reduces the errors by
$2.4\%,1.9\%,2.3\%$.
Adding the DESI\,$\times$\,PFS
cross-spectra, with their independent cross-stochastic
coefficients marginalized (C to D), gives a further
$1.4\%,0.7\%,1.0\%$ reduction.
The total incremental
PFS contribution (B to D) is therefore
$3.8\%,2.6\%,3.2\%$.
These fractional reductions have different denominators.

Among the three reported quantities, the growth error has
the largest fractional reduction, both from the DESI cross-spectra
and from the PFS addition.

The DESI cross-spectrum gain exceeds the
total PFS contribution most clearly for growth, while their
fractional effects on $\Mnu$ and $\Om$ are closer.
Along B--C--D, adding PFS auto-spectra gives a larger reduction
than subsequently adding the DESI\,$\times$\,PFS cross-spectra.

}

\begin{table}[t]
\centering\color{RevisionColor}
\small
\setlength{\tabcolsep}{5pt}
\begin{tabular}{lrrrrrr}
\hline
Case & $\sigma[f\se(z=1.1)]$ & $\Delta\%$ & $\sigma(\Mnu)$ [eV] & $\Delta\%$ & $\sigma(\Om)$ & $\Delta\%$ \\
\hline
A & 0.005894 & --- & 0.12155 & --- & 0.008432 & --- \\
B & 0.005478 & $-7.05$ & 0.11730 & $-3.49$ & 0.008115 & $-3.76$ \\
C & 0.005345 & $-2.43$ & 0.11511 & $-1.87$ & 0.007932 & $-2.25$ \\
D & 0.005271 & $-1.39$ & 0.11426 & $-0.74$ & 0.007854 & $-0.99$ \\
\hline
\end{tabular}
\caption{\color{RevisionColor}Fully marginalized $1\sigma$ uncertainties for the
cases shown in \cref{fig:full_area}, using the eight nominal
bins over $0.4\leq z\leq2.1$ and
$0.01\leq k\leq0.20\iMpch$. The growth quantity is derived
at $z=1.1$ from the global cosmological covariance.
Each $\Delta\%$ is $100(\sigma_{\mathrm{row}}/\sigma_{\mathrm{previous}}-1)$,
computed before rounding.}
\label{tab:headline}
\end{table}

\subsection{Information efficiency}
\label{sec:results_efficiency}

\begin{figure}[t]
\centering
\includegraphics[width=\textwidth]{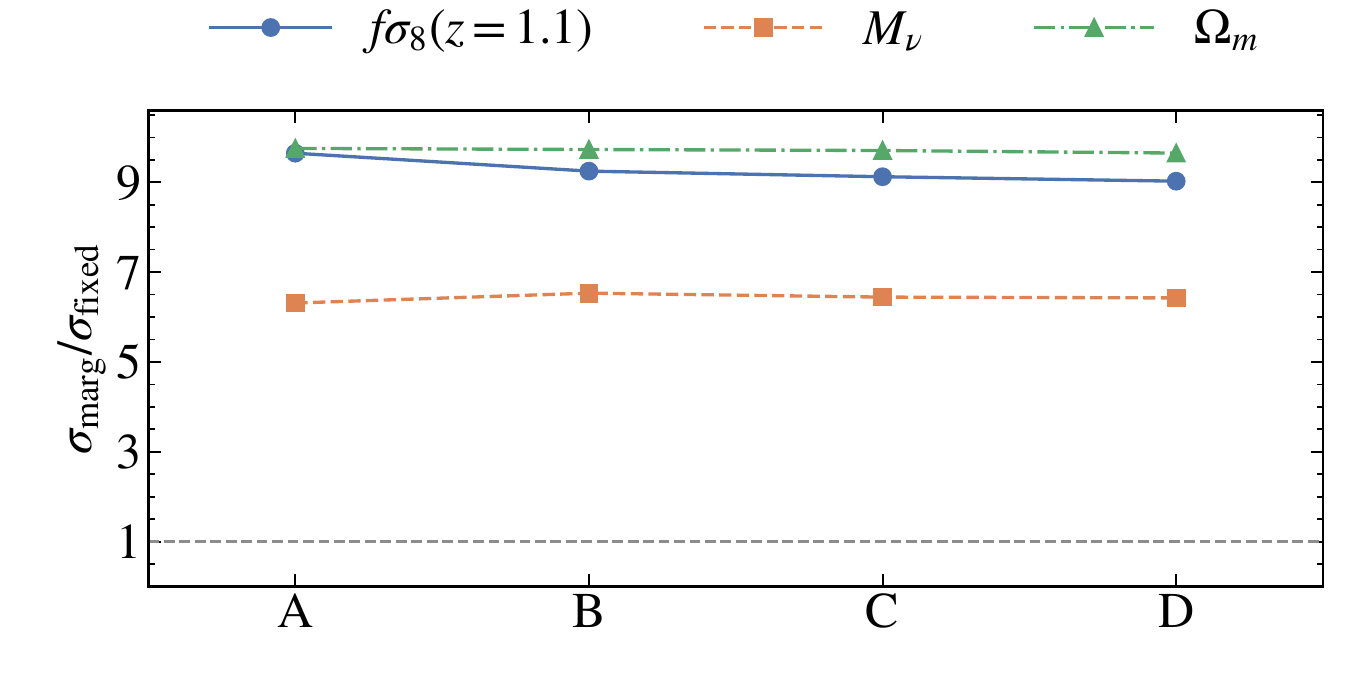}
\caption{\color{RevisionColor}Cost of nuisance marginalization for the matched full-survey
forecasts A/B/C/D at $\kmax=0.20\iMpch$.
The curves show $\mathcal D_p=\sigma_{\mathrm{marg}}(p)/
\sigma_{\mathrm{fixed}}(p)$ for $p=f\se(z=1.1)$, $\Mnu$ and $\Om$.
The dashed horizontal line marks unity, and the connecting lines guide the eye.
The numerator marginalizes over all nuisance coordinates with the stated priors,
and the denominator fixes them in the same basis (\cref{sec:results_efficiency}).
Other cosmological coordinates remain marginalized in both errors, with
$h,\omega_b,n_s$ fixed throughout.
The cases are defined in \cref{sec:scenarios} and shown in
\cref{fig:full_area}.}
\label{fig:efficiency}
\end{figure}

\revchange{\Cref{fig:efficiency} shows how nuisance uncertainty degrades
the cosmological precision in the four joint forecasts.
For each case we compare the marginalized errors with the
limit in which all its nuisance coordinates are fixed at
their fiducial values. Writing the total Fisher in blocks labeled by
the cosmological coordinates $\boldsymbol{\theta}_{\mathrm{cos}}$
and nuisance coordinates $\boldsymbol{\eta}$, the two
cosmological covariances are
\[
C_{\mathrm{cos}}^{\mathrm{marg}}
 =\left(F_{\mathrm{cos},\mathrm{cos}}
 -F_{\mathrm{cos},\eta}F_{\eta,\eta}^{-1}F_{\eta,\mathrm{cos}}\right)^{-1},
\qquad
C_{\mathrm{cos}}^{\mathrm{fixed}}=F_{\mathrm{cos},\mathrm{cos}}^{-1}.
\]
Here the blocks of $F$ include the priors specified in
\cref{sec:joint_fisher}. Both covariances
retain marginalization over the other cosmological
coordinates. We propagate each through the full
cosmological Jacobian to obtain the corresponding
$f\se(z=1.1)$ error.

We plot the marginalization penalty
$\mathcal D_p=\sigma_{\mathrm{marg}}(p)/\sigma_{\mathrm{fixed}}(p)$,
where $p$ denotes $f\se(z=1.1)$, $\Mnu$ or $\Om$.
The fixed limit refers to the coordinates in
\cref{tab:eft_params}, including $b_1\se$, the nonlinear
bias combinations, counterterms and stochastic coefficients
of every population and included cross-pair. Since these coordinates are amplitude-scaled, fixing $b_1\se$
while varying cosmology still lets $b_1$ vary.}

\revchange{\widowpenalty=10000\relax For B, C and D, respectively, the growth penalty is
$9.25,9.12,9.02$, the neutrino-mass penalty is
$6.53,6.44,6.42$, and the matter-density penalty is
$9.73,9.71,9.65$. Adding PFS lowers these penalties slightly.
From A to B, $\mathcal D_{\Mnu}$ increases
from $6.31$ to $6.53$, even though
$\sigma_{\mathrm{marg}}(\Mnu)$ decreases by $3.5\%$,
because $\sigma_{\mathrm{fixed}}(\Mnu)$ decreases faster,
by $6.7\%$.\par}

\clearpage
\subsection{\texorpdfstring{\revchange{Robustness of the gains}}{Robustness of the gains}}
\label{sec:results_sensitivity}

\revchange{We test whether the gains from adding DESI cross-spectra and PFS
depend on the analysis choices shared by the compared cases, or on
additional freedom in the spectrum model.}

\begin{figure}[t]
\centering
\includegraphics[width=\textwidth]{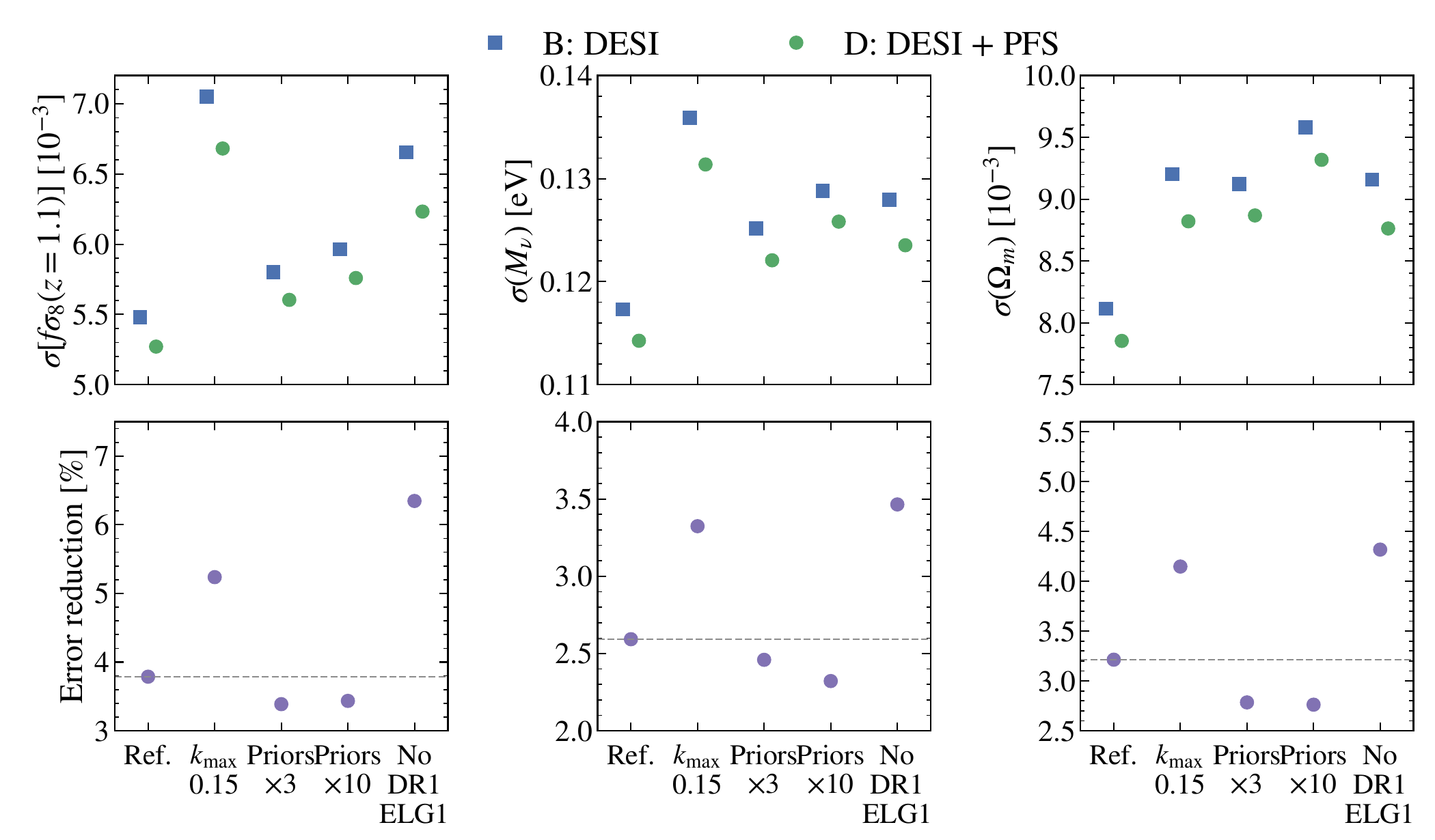}
\caption{\color{RevisionColor}Sensitivity of the joint forecasts to the common scale
cut, nuisance-prior widths and DESI DR1 ELG1 selection.
Upper panels: absolute marginalized errors for B
(DESI auto- and cross-spectra) and D (adding PFS autos
and DESI\,$\times$\,PFS crosses). The three columns show
the derived $f\se(z=1.1)$, $\Mnu$ and $\Om$.
Lower panels: positive error reductions
$100[1-\sigma_D/\sigma_B]$ for each matched comparison.
Dashed lines mark the reference values. The reference
uses $\kmax=0.20\iMpch$. Each alternative changes
one choice: the common cut to $0.15\iMpch$, all
finite nuisance-prior standard deviations by a factor
of three or ten, or exclusion of DESI DR1 ELG1 over
$0.8\leq z<1.1$. Flat nuisance priors, cosmological
priors and nominal redshift bins are unchanged.
The three-parameter cosmology and PFS baseline
$0.6\leq z\leq1.6$ are the same throughout.}
\label{fig:sensitivity}
\end{figure}

\revchange{\textbf{Scale cut, nuisance priors and sample selection.}
\Cref{fig:sensitivity} shows these tests. We
lower the common cut to $\kmax=0.15\iMpch$, multiply every finite
nuisance-prior width by three or ten, including those of the
cross-stochastic coefficients, or remove the DESI DR1 ELG1 interval
$0.8\leq z<1.1$ from both cases (all of ELG-1, the part of ELG-2 below
$z=1.1$, and their cross-spectra), as in the published DR1 full-shape
analyses~\cite{DESI:2024jxi,Chudaykin:2025aux}. Each change weakens the absolute constraints. With ten-times wider priors,
the errors in D on $f\se(z=1.1)$, $\Mnu$ and $\Om$ increase by $9.3/10.1/18.7\%$.
The B-to-D reductions change from $3.8/2.6/3.2\%$ in the reference to
$3.4/2.5/2.8\%$ and $3.4/2.3/2.8\%$ for three- and ten-times wider
priors, to $5.2/3.3/4.1\%$ at $\kmax=0.15\iMpch$, and to $6.3/3.5/4.3\%$
after removing DESI DR1 ELG1. The A-to-B reductions from DESI cross-spectra
grow with the prior width, from $7.1/3.5/3.8\%$ to $11.5/5.0/4.0\%$ and
$17.4/7.4/5.7\%$, since the cross-spectra matter more when the priors
constrain the nuisance parameters less. At $\kmax=0.15\iMpch$ they become
$3.9/4.3/4.9\%$.}

\revchange{\textbf{Cross-spectrum stochasticity.} These variants change the
stochastic model of the cross-spectra. \Cref{tab:crossmodel} reports the
C-to-D step, where the DESI\,$\times$\,PFS cross-spectra enter. Fixing their residual $\Pshot^{XY}$ and
$a_2^{XY}$ at zero roughly triples the growth gain, to $4.2\%$. The free stochastic coefficients
therefore absorb part of the cross-spectrum response, as discussed in
\cref{sec:drivers}. Every variant that keeps
these coefficients free leaves the C-to-D reduction in
$\sigma[f\se(z=1.1)]$ between $0.9\%$ and $1.8\%$. The variants are: flat priors on the coefficients, shared DESI/PFS ELGs with
$\nbar_{X\cap Y}=u\min(\nbar_X,\nbar_Y)$ (which relaxes the baseline
assumption of disjoint catalogs), nonzero fiducials for each cross pair, and
an added isotropic term
$a_{\mathrm{iso}}^{XY}(k/\knl)^2/\sqrt{\nbar_X\nbar_Y}$ in every included cross-spectrum, with an independent zero-fiducial coefficient for each pair
and bin.}

\begin{table}[t]
\centering\color{RevisionColor}
\small
\begin{tabular}{lrrr}
\hline
Variant & $f\se(z=1.1)$ & $\Mnu$ & $\Om$ \\
\hline
Reference & 1.39 & 0.74 & 0.99 \\
$\Pshot^{XY},a_2^{XY}$ fixed & 4.21 & 3.36 & 6.71 \\
$\Pshot^{XY},a_2^{XY}$ with flat priors & 1.09 & 0.49 & 0.54 \\
Shared ELGs, $u=1$ & 1.41 & 0.73 & 0.90 \\
Nonzero residual fiducials & 1.29--1.82 & 0.61--1.14 & 0.84--1.51 \\
$a_{\mathrm{iso}}^{XY}$, prior width 1 & 1.22 & 0.68 & 0.99 \\
$a_{\mathrm{iso}}^{XY}$, prior width 10 & 0.86 & 0.49 & 0.89 \\
$\mu^6$ FoG coefficient & 1.39--1.40 & 0.74--0.76 & 0.99--1.01 \\
\hline
\end{tabular}
\caption{\color{RevisionColor}C-to-D reductions (\%) of the marginalized errors for variants of the
cross-stochastic model and of the higher-order FoG term. Each variant changes
one assumption relative to the reference. The nonzero-fiducial row gives the
extremes over separate changes $\Pshot^{XY}=\pm0.5$ or $a_2^{XY}=\pm1$ of each
residual cross-pair coefficient, updating both mean and covariance. The $\mu^6$
FoG row spans prior widths of $100$, $400$ and
$1600\,[h_{\mathrm{ref}}^{-1}\mathrm{Mpc}]^4$ on $\mathcal A^2\ctilde_6^X$.}
\label{tab:crossmodel}
\end{table}

\revchange{\textbf{Additional FoG freedom.} In the higher-order FoG term of
\cref{sec:eft_model}, a single
$\ctilde^X$ ties the $\mu^4$ and $\mu^6$ field coefficients through $Z_X$.
Rubira and Conteddu~\cite{Rubira:2025scu} allow them to vary independently.
We test this freedom by adding
$-\ctilde_6^X k^4f^5\mu^6\delta_{\mathrm{lin}}/2$ to each tracer field,
so that it also enters the cross-spectra. The new coefficient has zero
fiducial, with an independent Gaussian prior on $\mathcal A^2\ctilde_6^X$ of width $100$,
$400$ or $1600\,[h_{\mathrm{ref}}^{-1}\mathrm{Mpc}]^4$. These widths are forecast assumptions. In A--D, the errors on $\ln(10^{10}A_s)$, $\Mnu$,
$\Om$ and $f\se(z=1.1)$ increase by less than $2\%$, and the incremental
error reductions change by less than $0.13$ percentage points.
The conclusions from A--D are therefore unchanged for these prior widths.}

\section{Discussion}
\label{sec:discussion}

\revchange{We interpret the gains in \cref{sec:results} using the bias--cosmology
correlations, and examine how the additional PFS information depends
on its redshift coverage, number density and bias. We then discuss how
multi-tracer constraints complement simulation-based priors.}

\Needspace{6\baselineskip}
\subsection{Which parameters drive the improvement?}
\label{sec:drivers}

\begin{figure}[t]
\centering
\includegraphics[width=\textwidth]{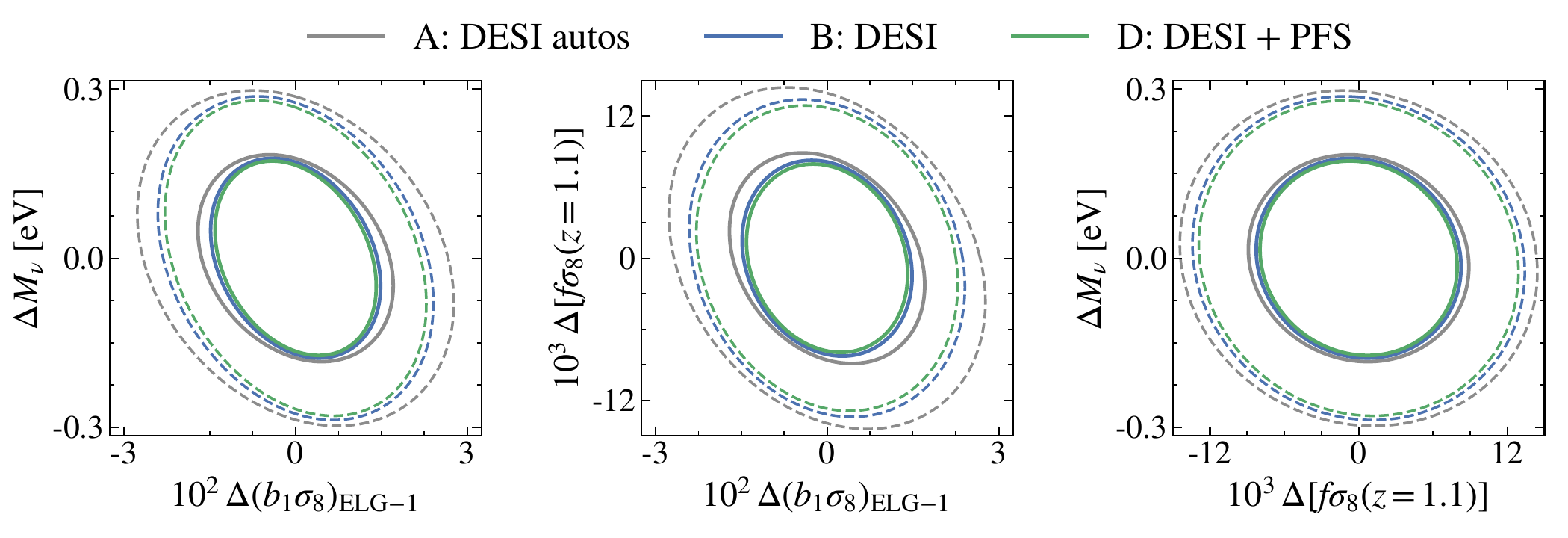}
\caption{\revchange{Marginalized 68\% (solid) and 95\% (dashed) joint
Gaussian contours} in the $b_1\se$--$\Mnu$,
$b_1\se$--$f\se$, and $f\se$--$\Mnu$ planes\revchange{.
Grey: A, DESI auto-spectra; blue: B,
adding DESI cross-spectra; green: D, adding PFS autos and DESI\,$\times$\,PFS cross-spectra.
All contours use the full inverse Fisher matrix, with
the stated priors and all other fitted coordinates
marginalized. The bias coordinate is ELG-1
$b_1\se$ in $0.8\leq z<1.0$, evaluated at
$z=0.9$. Growth is the derived $f\se(z=1.1)$,
propagated with its correlations with the bias and
$\Mnu$. Cosmology is shared across all eight bins.
Axes show deviations from the common fiducials,
$(b_1\se)_{\mathrm{fid}}=0.68223$,
$\Mnu^{\mathrm{fid}}=0.06\,\mathrm{eV}$, and
$[f\se(z=1.1)]_{\mathrm{fid}}=0.42065$. The comparison uses
the common $\kmax=0.20\iMpch$ and PFS baseline
$0.6\leq z\leq1.6$.}}
\label{fig:fisher_contours}
\end{figure}

\revchange{The linear bias alone does not account for the improvement from
cross-spectra.
Cross-spectra between two tracers in the same volume see the same
density field. In the linear, deterministic-bias and low-noise limit,
their relative biases can therefore be constrained beyond the usual sample-variance limit~\cite{Seljak:2008xr,McDonald:2008sh}. At one loop, the
information gain also depends on the nonlinear bias parameters
and tracer densities~\cite{Mergulhao:2023zso,Rubira:2025scu}.
\Cref{fig:fisher_contours}
shows the three parameter planes involving $b_1\se$,
$\Mnu$ and the derived $f\se$, with the bias shown for
ELG-1 in $0.8\leq z<1.0$. Its marginalized uncertainty
is 0.01131, 0.00984 and 0.00934 in A, B and D, respectively.
For the matched A--B comparison, adding DESI cross-spectra
reduces this uncertainty by 13.0\%, while adding PFS to B reduces
it by a further 5.1\%. The B--C and C--D steps give
1.0\% and 4.2\% reductions relative to their respective
starting cases. Thus, for this bias coordinate, most of the
additional PFS improvement occurs when its cross-spectra
are included. The corresponding B--D reductions in
$\sigma[f\se(z=1.1)]$ and $\sigma(\Mnu)$ are
3.8\% and 2.6\%. Better determination of this bias
therefore does not translate into an equally large gain
in cosmological precision.}

\revchange{The improvement does not require the bias to decorrelate from
cosmology. For ELG-1, $\rho(b_1\se,\Mnu)$ is $-0.276$, $-0.277$ and
$-0.289$ in B, C and D. Both errors decrease while this correlation
becomes slightly more negative. By contrast, $\rho[b_1\se,f\se(z=1.1)]$
becomes slightly less negative, from $-0.181$ to $-0.176$ and $-0.172$.
The cosmological and linear-bias correlation submatrices for every
redshift bin in A--D are provided as JCAP supplementary material, as
arXiv ancillary files and in the code repository.}

\revchange{With all nuisance coordinates fixed, the added DESI\,$\times$\,PFS
cross-spectra contribute $18$, $26$ and $38$ times more Fisher information on
$f\se(z=1.1)$, $\Mnu$ and $\Om$ than when the nuisance coordinates are
marginalized. Simultaneous shifts of bias, counterterm, FoG and stochastic
coefficients therefore absorb most of their cosmological response. The
fractional gain that remains comes mainly from constraining degenerate
nuisance coordinates. Fixing only the auto-spectrum stochastic coefficients
in both A and B lowers the A-to-B growth reduction from $7.1\%$ to $2.4\%$,
and fixing only $b_2\se^2$ and $b_{\G_2}\se^2$ lowers it to $3.0\%$.
Nuisance degeneracies also set much of the absolute error. Fixing only the
FoG coefficients in D reduces $\sigma(\Mnu)$ from $0.11426$ to
$0.09130\,\mathrm{eV}$.}

\Needspace{6\baselineskip}
\subsection{\texorpdfstring{\revchange{Redshift coverage, density and bias}}{Redshift coverage, density and bias}}
\label{sec:redshift}

\revchange{\textbf{Redshift coverage.} Extending PFS from $z=1.6$ to $z=2.4$ changes the total
growth-error reduction relative to B
from $3.8\%$ to $6.7\%$.
The corresponding reductions for $\Mnu$ and $\Om$ are
$6.7\%$ and $7.4\%$.
Adding the auto-spectra of PFS-ELG-6 and PFS-ELG-7 over
$1.6<z<2.1$ reduces the growth error by $1.7\%$. Including their
cross-spectra with QSO-5 and QSO-6, respectively, gives another
$0.5\%$, each relative to its preceding case. PFS-ELG-8, the PFS-only tail over
$2.1<z<2.4$, adds $0.9\%$.
The extension more than doubles the PFS gain in $\Mnu$ and $\Om$.

\textbf{Number density.} Starting from the forecast with PFS through
$z=2.4$, we raise the PFS density in one interval at a time, keeping its
clustering properties and stochastic-power priors fixed. Doubling the
density reduces the errors on $\Mnu$ and $\Om$ by at most $1.0\%$, in
PFS-ELG-8 ($0.9\%$ in PFS-ELG-6), and the growth error by at most
$0.6\%$, in PFS-ELG-6. Below $z=1.6$, each doubling reduces these errors
by less than $0.3\%$. However, the intervals contain very different
numbers of galaxies: doubling PFS-ELG-6 at $1.6<z<2.0$ adds about
$566{,}000$ galaxies, compared with $85{,}000$ for PFS-ELG-1 at
$0.6<z<0.8$. Per added galaxy, a small density increase in PFS-ELG-1
reduces the growth error more than twice as much as one in PFS-ELG-6,
while PFS-ELG-7 and PFS-ELG-8 above $z=2$ are the most efficient for
$\Mnu$ and $\Om$.

\textbf{Linear bias.} In the same forecast, at fixed density, we multiply
the linear bias of PFS-ELG-1, PFS-ELG-5 or PFS-ELG-6 by factors from $0.8$
to $2.1$. We either change $b_1$ alone or also recompute the fiducial $b_2$,
$b_{\G_2}$, $b_{\Gamma_3}$ and $c_2$ from the new $b_1$ with the relations of
\cref{sec:eft_fiducials}. All nuisance parameters remain free, with unchanged
priors. The errors on
$f\se(z=1.1)$, $\Mnu$ and $\Om$ change by at most $0.4\%$, $1.9\%$ and
$2.4\%$, respectively. The largest reductions, in $\Mnu$ and $\Om$, come
from doubling the bias of PFS-ELG-6 at $1.6<z<2.0$. Within this model,
extending the PFS redshift coverage therefore reduces the errors more than
doubling the density or changing the bias of any single PFS interval.}

\Needspace{5\baselineskip}
\subsection{Complementarity with simulation-based priors}

\revchange{Chudaykin et al.~\cite{Chudaykin:2026nls} report that their
power-spectrum analysis using SBPs yields $b_1$ posterior
widths about $30\%$ smaller for BGS/LRG and $40\%$
smaller for DESI DR1 ELG2/QSO than their
power-spectrum-plus-bispectrum analysis with conservative EFT priors.
The multi-tracer constraints use only the clustering likelihood
and the population-uniformity assumption in \cref{sec:hod_uniformity}.
In D, this likelihood constrains $b_1\se$ to $0.007$--$0.028$ and narrows
$b_2\se^2$ and $b_{\G_2}\se^2$ to less than $0.2$ of their prior widths in at
least 18 of the 19 tracer--bin sets, while $\mathcal A b_{\Gamma_3}$,
$\mathcal A c_2$ and $a_2$ keep median posterior-to-prior ratios of $0.92$,
$0.57$ and $0.70$ (\cref{fig:calibrated_vs_broad}). SBPs can therefore add the
most information on $b_{\Gamma_3}$, the counterterms and the $k^2$ stochastic terms.}

\revchange{This distinction motivates a cross-check of the
prior dependence of amplitude constraints, including those
relevant to $S_8$. In App.~D of
Chudaykin et al.~\cite{Chudaykin:2026nls}, replacing only the quasar
SBPs by conservative priors shifts $\se$ from
$0.766^{+0.015}_{-0.016}$ to
$0.786^{+0.024}_{-0.026}$ in the
power-spectrum-plus-BAO analysis. A future DESI\,$\times$\,PFS posterior analysis could test such prior sensitivity
through additional clustering information.}

The method generalizes to any pair of overlapping spectroscopic
surveys with different target
selections~\cite{Alarcon:2016bkr}---Euclid$\,\times\,$DESI,
Roman$\,\times\,$Spec-S5, and
DESI-II$\,\times\,$MegaMapper---and joint
J-PAS\,$\times$\,PFS forecasts have already explored similar
synergies~\cite{Qin:2025nkk}. Adding the tree-level
bispectrum~\cite{Ivanov:2023qzb,Bakx:2025pop} would activate
the sub-leading counterterm $c_1$ and break the
$b_1$--$b_2$--$b_{\G_2}$ degeneracies through the $Z_2$
kernel structure.

The Gaussian Fisher approximation captures the leading
information content but not non-Gaussian posterior structure\revchange{. An} MCMC analysis with the full likelihood would sharpen the
comparison with SBPs.

The PFS-ELG EFT fiducials are scaled from
DESI rather than fitted to PFS mocks\revchange{. Validation against upcoming}
mock suites---Uchuu-LRG~\cite{FernandezGarcia:2025rso},
Uchuu-ELG~\cite{Vaisakh:inprep}, and Mucho
Uchuu~\cite{Ishiyama:inprep} on the DESI side, and PFS galaxy
mocks~\cite{PFS-CO-mock:inprep} on the PFS side---will anchor
the EFT fiducials and \revchange{test the population-uniformity assumption}
(\cref{sec:hod_uniformity}).

\section{Conclusions}
\label{sec:conclusions}

\revchange{Mergulh\~{a}o et al.~\cite{Mergulhao:2021kip,Mergulhao:2023zso}
established one-loop multi-tracer gains in real and redshift space, and
Rubira and Conteddu~\cite{Rubira:2025scu} forecast DESI and PFS separately,
each at one effective redshift. Here we combine both surveys over eight
redshift bins and the partitioned footprint, and quantify what the
overlapping PFS sample adds once the internal DESI multi-tracer information
is included. The main comparison uses matched redshift bins
and priors, PFS over $z\in[0.6,1.6]$, and a common fit of
$\ln(10^{10}A_s),\Mnu,\Om$, with $h,\omega_b,n_s$ fixed and
correlated $f\se(z)$ predictions:}

\begin{enumerate}
\item \revchange{The volume-partitioned Fisher analysis}
  \[\revchange{F}=F_{\mathrm{overlap}}(1{,}200\,\mathrm{deg}^2)
   + F_{\mathrm{nonoverlap}}(12{,}800\,\mathrm{deg}^2)\]
  combines the internal-DESI channel over the full footprint with
  \revchange{PFS auto-spectra and the DESI\,$\times$\,PFS cross-spectra} in the
  $1{,}200\;\mathrm{deg}^2$ intersection, \revchange{within the independent-volume
  approximation.}

\item \revchange{Adding DESI cross-spectra (A to B) reduces the errors on the derived $f\se(z=1.1)$,
  $\Mnu$ and $\Om$ by $7.1\%$, $3.5\%$ and
  $3.8\%$, respectively. The DESI multi-tracer analysis (B) reaches
  $\sigma[f\se(z=1.1)]=0.005478$,
  $\sigma(\Mnu)=0.1173\,\mathrm{eV}$, and
  $\sigma(\Om)=0.008115$} at $\kmax=0.20\iMpch$ without any
  cross-survey input.

\item Adding PFS to the $1{,}200\;\mathrm{deg}^2$ overlap further
  tightens \revchange{$\sigma[f\se(z=1.1)]$ by $3.8\%$,
  $\sigma(\Mnu)$ by $2.6\%$, and $\sigma(\Om)$ by
  $3.2\%$ (B to D), reaching
  $\sigma[f\se(z=1.1)]=0.005271$,
  $\sigma(\Mnu)=0.1143\,\mathrm{eV}$, and
  $\sigma(\Om)=0.007854$. Both the PFS auto-spectra and the DESI\,$\times$\,PFS cross-spectra contribute. In the main comparison, the largest fractional gain among these
  quantities is on growth. The ELG-1 $b_1\se$ error in
  $0.8\leq z<1.0$ decreases by $5.1\%$, but this does not
  uniquely identify linear bias as the cause of the cosmological
  gains. Extending PFS to $z=2.4$ increases the reductions in
  $\sigma[f\se(z=1.1)]$, $\sigma(\Mnu)$ and $\sigma(\Om)$ relative to B to
  $6.7\%$, $6.7\%$ and $7.4\%$, respectively. Within this model, this
  extension reduces the errors more than doubling the density or changing
  the bias of any single PFS interval.}

\item \revchange{The joint analysis complements simulation-based priors.
  Multi-tracer auto- and cross-spectra constrain the linear and quadratic
  biases and the shot noise jointly with cosmology, while $b_{\Gamma_3}$, the
  counterterms and the $k^2$ stochastic terms stay closer to their priors.
  For these, SBPs supply external information from simulated galaxy--halo
  connections. A future joint posterior analysis could test the prior sensitivity of
  SBP-based amplitude constraints relevant to $S_8$.}

\item \revchange{Under the same independent-volume approximation, the framework
  extends} to \revchange{$N$} overlapping spectroscopic surveys. For \revchange{$N$} footprints, the union partitions
  into up to $2^N{-}1$ disjoint regions, one per non-empty
  subset of surveys\revchange{. Each} region carries a multi-tracer Fisher
  built from whichever tracers are active there, and all region
  Fishers add in a common parameter space (cosmology shared
  globally, nuisance unique per tracer, per redshift bin). This
  accommodates upcoming multi-survey configurations
  such as \revchange{DESI\,$\times$\,PFS\,$\times$\,Euclid\,$\times$\,Roman} in
  partial overlaps. \revchange{Joint J-PAS\,$\times$\,PFS forecasts have already been
  studied~\cite{Qin:2025nkk}.}
\end{enumerate}

\acknowledgments
I thank Kazuyuki Akitsu for engaging discussions about simulation-based priors, Haruki Ebina for helpful comments about tracer cross-stochasticity, as well as David Piras and \revchange{Hidde} Jense for useful correspondences about \texttt{cosmopower-jax} and the \texttt{CosmoPower} emulator.

I am grateful to Linda Blot and Jingjing Shi for heartening and uplifting conversations between the two VJA red-eye flights on which this work was begun and finished.

It takes two villages to build the PFS and DESI surveys; I thank everyone in the PFS and DESI collaborations, whose collective efforts made these surveys possible and successful.
I acknowledge support from the Japan Foundation for Promotion of Astronomy Research Grant and the JSPS KAKENHI Grant Numbers 25K23373 and 26H00404.
This work was supported by World Premier International Research Center Initiative (WPI Initiative), MEXT, Japan.

This study utilized the following open-source libraries and software packages:
\href{https://github.com/google/jax}{\code{JAX}},
\href{https://matplotlib.org/}{\code{Matplotlib}},
\href{https://numdifftools.readthedocs.io/en/master/}{\code{numdifftools}},
\href{https://numpy.org/}{\code{NumPy}},
\href{https://pyyaml.org/}{\code{PyYAML}},
\href{https://docs.pytest.org/}{\code{pytest}},
\href{https://scipy.org/}{\code{SciPy}},
\href{https://www.sympy.org/}{\code{SymPy}},
\href{https://github.com/f0uriest/quadax}{\code{quadax}},
and \href{https://github.com/f0uriest/interpax}{\code{interpax}}.

\appendix

\section{Validation of the Fisher pipeline}
\label{app:Fisher_convergence}

We validate the ingredients of our Fisher forecast against a combination of numerical convergence tests and analytical limit checks.

\subsection{Jacobian matrix}
\label{app:derivatives}

\revchange{All power spectrum derivatives $\partial P_\ell/\partial\theta_i$,
where $\ell$ labels the multipole order and $i$ a varied cosmological
or nuisance parameter,} are computed via JAX forward-mode automatic differentiation through the
\revchange{one-loop \texttt{ELF} model, including the intrinsic response and the
Alcock--Paczy\'nski mapping of the complete mean. We compare them with
symmetric finite differences at three successively halved displacements.
The comparison uses a covariance-weighted norm over the retained
multipoles and wavenumbers, with a relative tolerance of $10^{-3}$
at two successive steps. Analytically vanishing responses are checked
against an absolute tolerance relative to the fiducial signal.}

\begin{figure}[t]
\centering
\includegraphics[width=\textwidth]{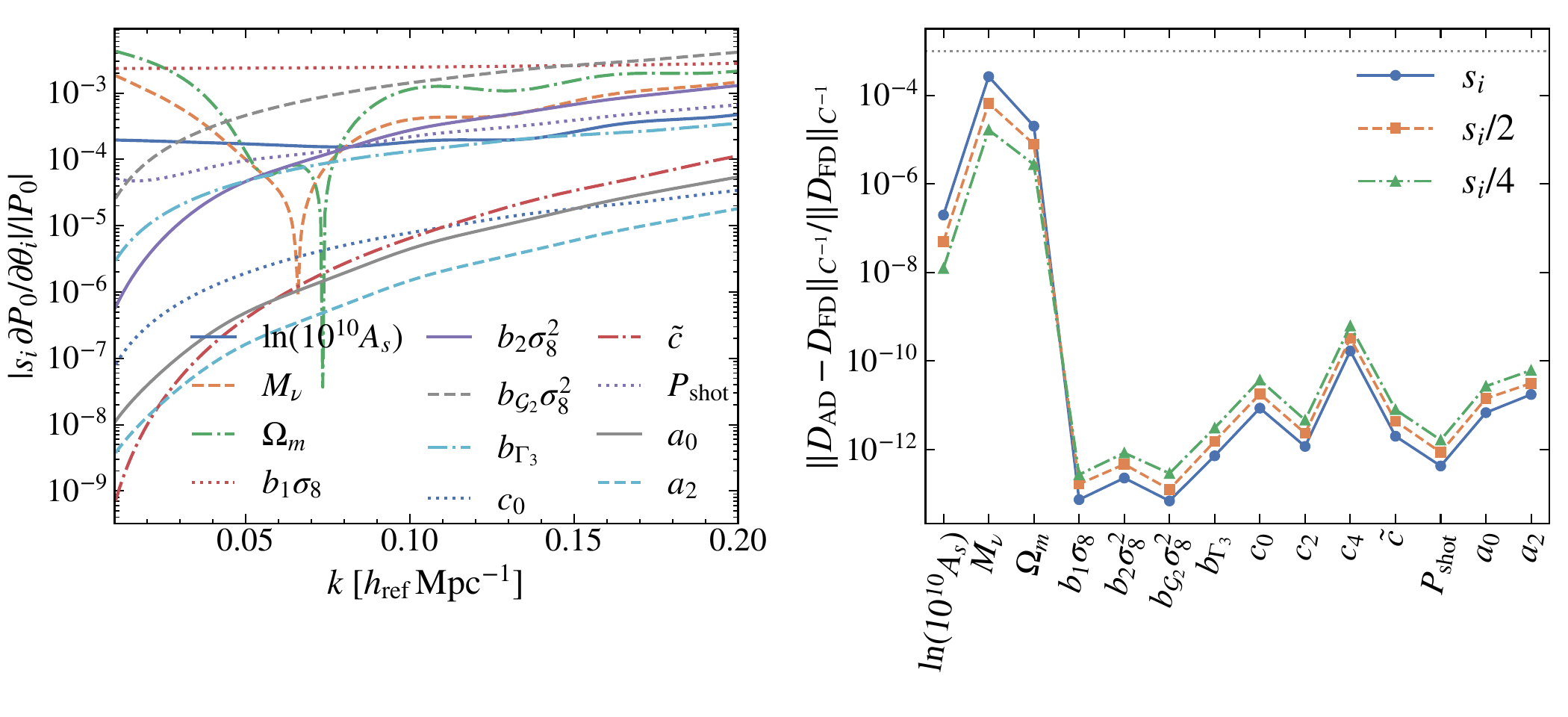}
\caption{\revchange{Numerical derivative check for the ELG-1 auto-spectrum at
$z=0.9$, using $0.01<k<0.20\iMpch$.
Left: $|s_i\partial P_0/\partial\theta_i|/|P_0|$, where $i$ labels
one of the three cosmological or 11 nuisance coordinates and $s_i$
is its displacement. The nuisance labels suppress the amplitude factors
defined in \cref{tab:eft_params}.
The largest tested displacement is $s_i=10^{-3}$ in the
parameter units of \cref{tab:eft_params}, except
$s_{\Mnu}=0.005\,\mathrm{eV}$,
$s_{c_2}=0.03\,(\Mpch)^2$, and
$s_{\ctilde}=0.4\,(\Mpch)^4$.
The vanishing $c_2$ and $c_4$ monopole responses are omitted
from the logarithmic panel. Right: covariance-weighted relative
AD--finite-difference discrepancies over $\ell=0,2,4$
for steps $s_i,s_i/2,s_i/4$. The dotted line marks $10^{-3}$.}}
\label{fig:deriv_validation}
\end{figure}

\revchange{\Cref{fig:deriv_validation} shows the power-spectrum derivatives and
their numerical comparison for the ELG-1 auto-spectrum at $z=0.9$.
The local parameter set contains the three cosmological coordinates
$\{\ln(10^{10}A_s),\Mnu,\Om\}$ and 11 active nuisance
coordinates. The left panel shows dimensionless monopole responses to
the stated parameter displacements. The right panel uses
all three multipoles. Its maximum relative differences are
$2.7\times10^{-4}$, $6.7\times10^{-5}$, and
$1.7\times10^{-5}$ as the step is halved.}

\subsection{Covariance matrix}
\label{app:covariance}

The Gaussian multipole covariance for $N_t$ tracers
follows from Wick-contracting two power spectrum
estimators:
\be
\label{eqn:cov_wick}
\Cov\!\bigl[P_\ell^{XY},\, P_{\ell'}^{WZ}\bigr]\!(k)
= \frac{(2\ell{+}1)(2\ell'{+}1)}{2\,N_{\mathrm{modes}}(k)}
  \!\int_{-1}^{1}\!\dd\mu\; L_\ell(\mu)\, L_{\ell'}(\mu)\,
  \bigl[
    P^{XW}\! P^{YZ}
  + P^{XZ}\! P^{YW}
  \bigr],
\ee
where $N_{\mathrm{modes}}(k)=k^2\Delta k\,V/(2\pi^2)$ is the number
of Fourier modes in a shell of width $\Delta k$, and
$L_\ell(\mu)$ are the Legendre polynomials. \revchange{The labels
$X,Y,W,Z$ run over the $N_t$ tracer populations and $\ell,\ell'$
over the retained multipole orders $0,2,4$.} The two terms inside
the brackets are the two Wick pairings of the 4 \revchange{tracer} fields\revchange{. For} a single tracer ($X{=}Y{=}W{=}Z$) they are identical,
recovering the familiar factor-of-two scaling of the power-spectrum
variance.

The power $P^{XY}(k,\mu)$ entering the covariance includes the
\revchange{one-loop model and all its stochastic terms, with Poisson noise
added once:}
\be
\label{eqn:ptot}
P^{XY}(k,\mu) =
  \revchange{P^{XY,\mathrm{model}}(k,\mu)}
  {+\frac{\delta_{XY}^{\mathrm{K}}}{\nbar_X}\revchange{\,.}}
\ee
\revchange{Here $\delta_{XY}^{\mathrm{K}}$ is the Kronecker delta, so the
Poisson term contributes only to the auto-spectra.
The fiducial non-Poisson stochastic coefficients vanish, while their
responses remain in the mean derivatives, including the independent
cross-stochastic coordinates in \cref{eqn:cross}.}

\begin{figure}[t]
\centering
\includegraphics[width=\textwidth]{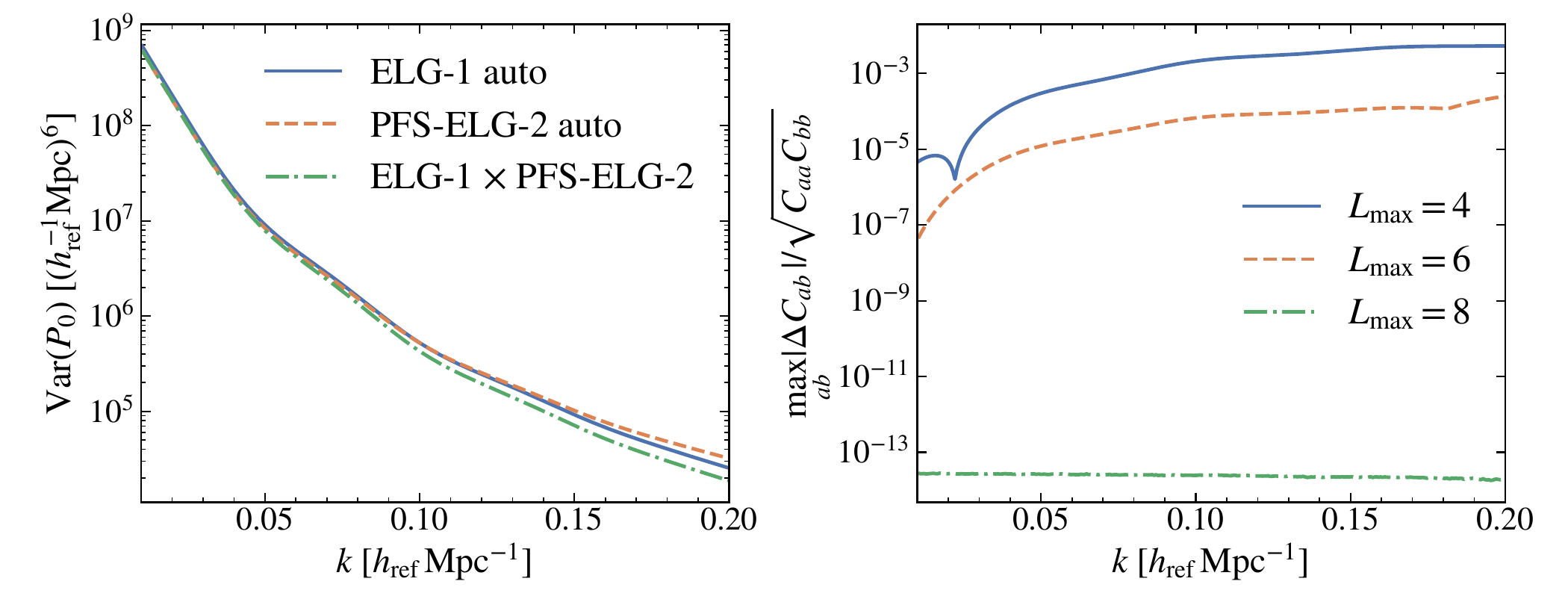}
\caption{\revchange{Numerical check of the Gaussian covariance at $z=0.9$.
Left: monopole variances for ELG-1, PFS-ELG-2
and their cross-spectrum in the $1{,}200\,\mathrm{deg}^2$
overlap, with $\Delta k=0.000625\iMpch$.
Right: maximum $|\Delta C_{ab}|/\sqrt{C_{aa}C_{bb}}$ over the
nine observables indexed by $a,b$, formed from these three spectra and
$\ell=0,2,4$, comparing 64-point Gauss--Legendre quadrature
with the Wigner~3$j$ calculation using spectrum multipoles
through $L_{\mathrm{max}}=4,6,8$.}}
\label{fig:cov_validation}
\end{figure}

At each wavenumber, the observable vector has
\revchange{$3N_{\mathrm{spec}}$ elements for the monopole, quadrupole and hexadecapole,
$\ell=0,2,4$.} The
$\mu$-integrals are evaluated with \revchange{64}-point GL
quadrature. An equivalent approach is the analytic Wigner~3$j$ decomposition of Rubira \revchange{and}
Conteddu~\cite{Rubira:2025scu} (their Eq.~2.20), which first decomposes $P(k,\mu)$ into multipoles $P_\ell$ and then evaluates the products of Legendre polynomials via
$L_{\ell_1}L_{\ell_2}=\sum_{\ell_3}(2\ell_3{+}1)
\bigl(\begin{smallmatrix}\ell_1 & \ell_2 & \ell_3 \\
0 & 0 & 0\end{smallmatrix}\bigr)^{\!2} L_{\ell_3}$.
\revchange{Here $\ell_1,\ell_2$ are the orders of the two Legendre factors.
The sum runs over $|\ell_1-\ell_2|\leq\ell_3\leq\ell_1+\ell_2$
with $\ell_1+\ell_2+\ell_3$ even.
The} GL quadrature integrates the full $P(k,\mu)$ directly, while the
Wigner~3$j$ formula truncates at the maximum multipole included in the
\revchange{spectrum expansion. This truncation is distinct from retaining only
$\ell=0,2,4$ in the observable vector, since higher multipoles of the
full spectrum can contribute to their covariance.
For the fiducial ELG-1 and PFS-ELG-2 spectra at $z=0.9$,
\cref{fig:cov_validation} gives maximum normalized covariance
residuals of $5.3\times10^{-3}$, $2.5\times10^{-4}$, and
$2.8\times10^{-14}$ when the spectrum expansion ends at
$L_{\mathrm{max}}=4,6,8$, respectively. The normalization is
$|\Delta C_{ab}|/\sqrt{C_{aa}C_{bb}}$, where $a,b$ index the
spectrum multipoles at fixed $k$. Here $C$ is the GL covariance
and $\Delta C$ the difference between the two evaluations.}

\subsection{Fisher matrix}
\label{app:fisher_checks}

\begin{figure}[t]
\centering
\includegraphics[width=0.75\textwidth]{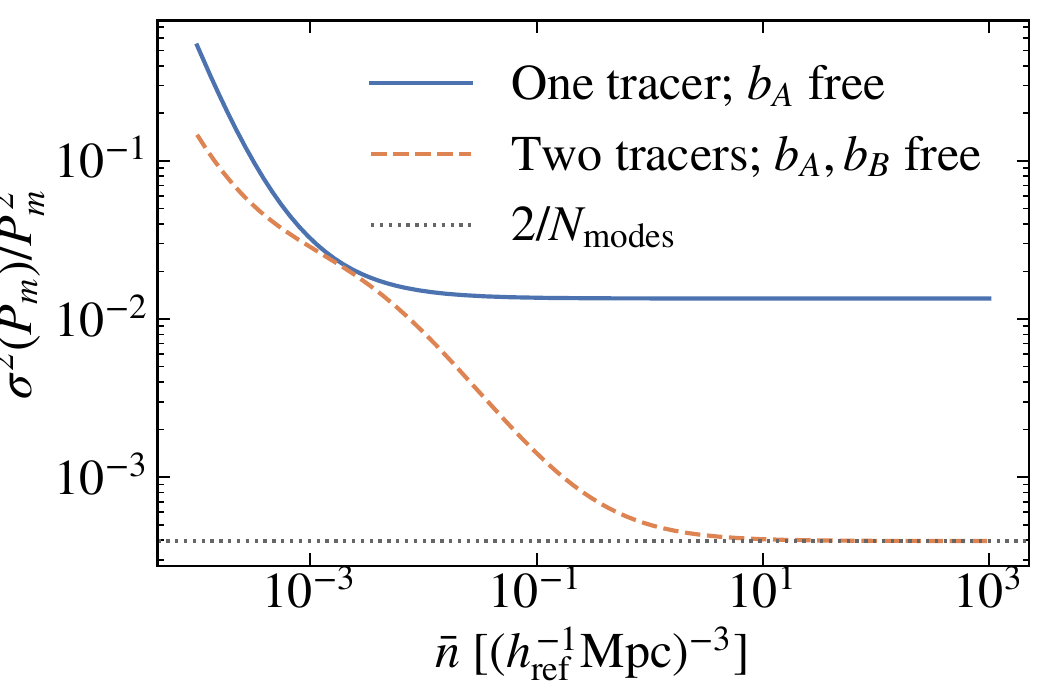}
\caption{\revchange{Recovery of the matter-amplitude sample-variance limit in the
linear Kaiser model. Solid: one tracer with $b_A$ marginalized;
dashed: two tracers with $b_A,b_B$ marginalized. The dotted line is
$2/N_{\mathrm{modes}}$.} Parameters: $b_A=1$, $b_B=2$\revchange{,
fixed} $f=0.8$, $P_m=10^3\,(\Mpch)^3$\revchange{,}
$k=0.1\iMpch$, $\Delta k=0.01\iMpch$\revchange{, and}
$V=10^9\,(\Mpch)^3$\revchange{, with equal number densities and
independent Poisson noise.}}
\label{fig:ms09_convergence}
\end{figure}

We verify the assembled Fisher matrix against four consistency
checks and one limit recovery:

\begin{enumerate}
\revchange{\item \textbf{Analytic normalization.}
  In the isotropic single-tracer limit, direct evaluation recovers
  $\operatorname{Var}P_0=2(P+1/\nbar)^2/N_{\mathrm{modes}}$.
  The shell width enters once through $N_{\mathrm{modes}}$ in the
  Fisher sum over wavenumber shells (\cref{sec:mt_fisher}).}

\revchange{\item \textbf{Information monotonicity.}
  With the same covariance model and shared parameters, retaining
  additional observables cannot decrease Fisher information. We check
  this on a common parameter space for $A\to B\to C\to D$.}

\item \textbf{Volume partition identity.}
  \revchange{Within the Gaussian approximation with independent disjoint survey
  regions, the Fisher matrix is linear in volume for a
  fixed wavenumber grid, number density and mean model. The independence
  assumption omits correlations between regions from shared
  long-wavelength modes through SSC~\cite{Wadekar:2019rdu,Sugiyama:2019ike}.
  For the
  ELG-1 auto-spectrum in $[0.8,1.0]$, the sum of the saved
  overlap and non-overlap contributions agrees with evaluation at
  their total volume to better than $10^{-12}$ in relative matrix
  norm. Their areas are $1{,}200$, $12{,}800$, and
  $14{,}000\,\mathrm{deg}^2$, respectively. This checks}
  $F(V_{\mathrm{overlap}})+F(V_{\mathrm{nonoverlap}})=F(V_{\mathrm{full}})$
  \revchange{under the stated approximation.}

\revchange{\item \textbf{Case ordering.}
  With the same priors on shared parameters, the marginalized
  cosmological covariance decreases in the positive-semidefinite
  sense for $A\to B$, $B\to C$, $C\to D$, and
  $B\to D$.}

\Needspace{10\baselineskip}
\revchange{\item \textbf{Linear multi-tracer limit.}
  In the Kaiser model for two tracers labeled $A,B$, with $b_A=1$, $b_B=2$, fixed
  $f=0.8$, independent Poisson noise and no counterterms,
  we marginalize over both biases. The fractional variance of the
  common matter power $P_m$ approaches}
  \be
  \label{eqn:ms09_floor}
  \frac{\sigma^2(P_m)}{P_m^2}
  \;\xrightarrow[\nbar\to\infty]{}\;
  \frac{2}{N_{\mathrm{modes}}}\,.
  \ee
  \revchange{This is the usual sample-variance limit for the matter amplitude,
  equivalently for $f^2P_m$ when $f$ is fixed. It is the
  ratios of tracer responses whose errors can fall below this
  sample-variance floor~\cite{Seljak:2008xr,McDonald:2008sh}.
  \Cref{fig:ms09_convergence} approaches the amplitude limit to
  $0.03\%$ at $\nbar P_m=10^6$. The single-tracer
  result remains above it because the unknown bias must be inferred
  from the angular dependence. The mode count includes both
  $\boldsymbol{k}$ and $-\boldsymbol{k}$, with the usual
  Gaussian factor $1/2$.}
\end{enumerate}

\revchange{For A--D we compare the cosmological covariances obtained
from the full inverse Fisher matrix, the Schur expression in
\cref{sec:results_efficiency} and independent augmented least squares,
with the same priors throughout. Their differences satisfy
$\max_{ij}|\Delta C_{ij}|/\sqrt{C_{ii}C_{jj}}<3\times10^{-11}$.
Evaluating the likelihood at finite displacements along $f\se(z=1.1)$,
$\Mnu$ and $\Om$, with the nuisance parameters either fixed or shifted by
the compensation that the Fisher matrix predicts, reproduces the Fisher
curvature to relative $10^{-3}$ at two step sizes in all four cases.}

\begin{figure}[t]
\centering
\includegraphics[width=\textwidth]{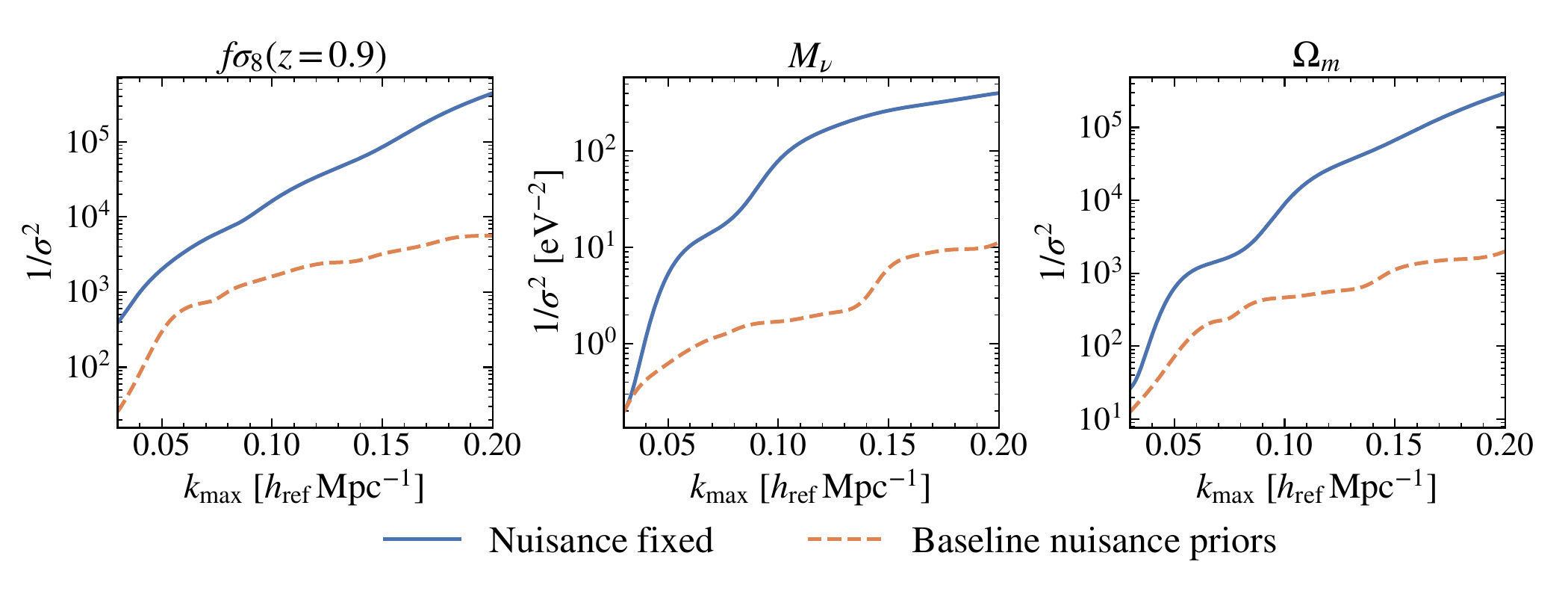}
\caption{\revchange{Scale dependence of accumulated constraints from ELG-1
auto-multipoles in $0.8<z<1.0$, using the full
$14{,}000\,\mathrm{deg}^2$ area. The inverse variances include
all shells from $k_{\mathrm{min}}=0.01\iMpch$ up to the displayed
$k_{\mathrm{max}}$. Solid: nuisance parameters fixed; dashed:
nuisance parameters marginalized with the baseline nuisance priors.
Other cosmological coordinates are marginalized in both curves,
with $h,\omega_b,n_s$ fixed and $f\se(z=0.9)$ derived from
$\{\ln(10^{10}A_s),\Mnu,\Om\}$.}}
\label{fig:fisher_info}
\end{figure}

\revchange{\Cref{fig:fisher_info} shows how inverse variances accumulate as
$k_{\mathrm{max}}$ increases for the ELG-1 auto-spectrum at $z=0.9$
over the full DESI area. At each upper cut we sum the likelihood
Fisher matrices of the retained shells and add the priors once.
The curves compare fixed nuisance parameters with marginalization
over the 11 active nuisance coordinates under the baseline nuisance
priors. The other cosmological coordinates are marginalized in both
cases, and $f\se(z=0.9)$ is derived from the global cosmology. At
$k_{\mathrm{max}}=0.20\iMpch$, the nuisance-induced variance ratios
are approximately $79$, $36$, and $147$ for $f\se(z=0.9)$,
$\Mnu$, and $\Om$, respectively.}

\Needspace{8\baselineskip}
\section{EFT priors from co-evolution}
\label{app:coevolution_priors}

In the baseline analysis,
the nonlinear biases $b_2$ and $b_{\G_2}$ are assigned independent
Gaussian priors ($\sigma=5$ for $b_2\se^2$ and $b_{\G_2}\se^2$).
\revchange{The fiducial $b_2$ follows the fit of
Lazeyras et al.~\cite{Lazeyras:2015lgp}, and the tidal bias uses
the co-evolution prescription in \cref{sec:eft_fiducials}.}

\revchange{Mergulh\~{a}o et al.~\cite{Mergulhao:2021kip} examined Gaussian
priors centered on bias co-evolution relations, with finite scatter,
in their real-space multi-tracer power-spectrum analysis
(Appendix~B). The improvement was small in that test.}

\revchange{A co-evolution prior would correlate the nonlinear biases with
$b_1$. We do not impose such priors in the forecasts reported here.}

\clearpage
\bibliographystyle{JHEP}
\bibliography{PFSxDESI-multi-fish}

\end{document}